\documentclass{pasj01}
\draft 
\Received{$\langle$reception date$\rangle$}
\Accepted{$\langle$acception date$\rangle$}
\Published{$\langle$publication date$\rangle$}
\usepackage{color}

\begin{document}

\title{Cluster formation induced by a cloud--cloud collision in [DBS2003]179}
\author{Sho Kuwahara$^{1}$}
\author{Kazufumi Torii$^{2}$}
\author{Norikazu Mizuno$^{3}$}
\author{Shinji Fujita$^{4*}$}
\author{Mikito Kohno$^{4}$}
\author{Yasuo Fukui$^{4}$}

\altaffiltext{1}{Department of Astronomy, School of Science, The University of Tokyo, 7-3-1 Hongo, Bunkyo-ku, Tokyo, Japan 133-0033}
\altaffiltext{2}{Nobeyama Radio Observatory, 462-2 Minamimaki, Minamisaku, Nagano, Japan 384-1305}
\altaffiltext{3}{National Astronomical Observatory of Japan, Mitaka, Tokyo, Japan 181-8588}
\altaffiltext{4}{Department of Astrophysics, Nagoya University, Furo-cho, Chikusa-ku, Nagoya, Aichi, Japan 464-8602}

\email{fujita.shinji@a.phys.nagoya-u.ac.jp}



\maketitle

\begin{abstract}
[DBS2003]179 is a super star cluster in the Galaxy discovered by deep near infrared observations. We carried out CO $J$=1--0 and $J$=3--2 observations of the region of  [DBS2003]179 with NANTEN2, ASTE and the Mopra 22m telescope. We identified and mapped two molecular clouds which are likely associated with the cluster. The association is evidenced by the spatial correlation with the 8\,$\mu$m {\it Spitzer} image, and a high ratio of the two transitions of $^{12}$CO ($J$=3--2 to $J$=1--0). The two clouds show complementary distribution in space and bridge features connecting them in velocity. We frame a hypothesis that the two clouds collided with each other 1--2\,Myrs ago, and the collision compressed the interface layer, triggering the formation of the cluster. This offers an additional piece of evidence for a super star cluster formed by cloud--cloud collision alongside of the four super star clusters including Wd2, NGC3603, RCW38 and R136. The findings indicate that the known super star clusters having closely associated dust emission are formed by cloud--cloud collision, lending support for the important role of cloud--cloud collision in high-mass star formation.
\end{abstract}

\section{Introduction}
\begin{figure}
  \begin{center}
  \includegraphics[width=14cm]{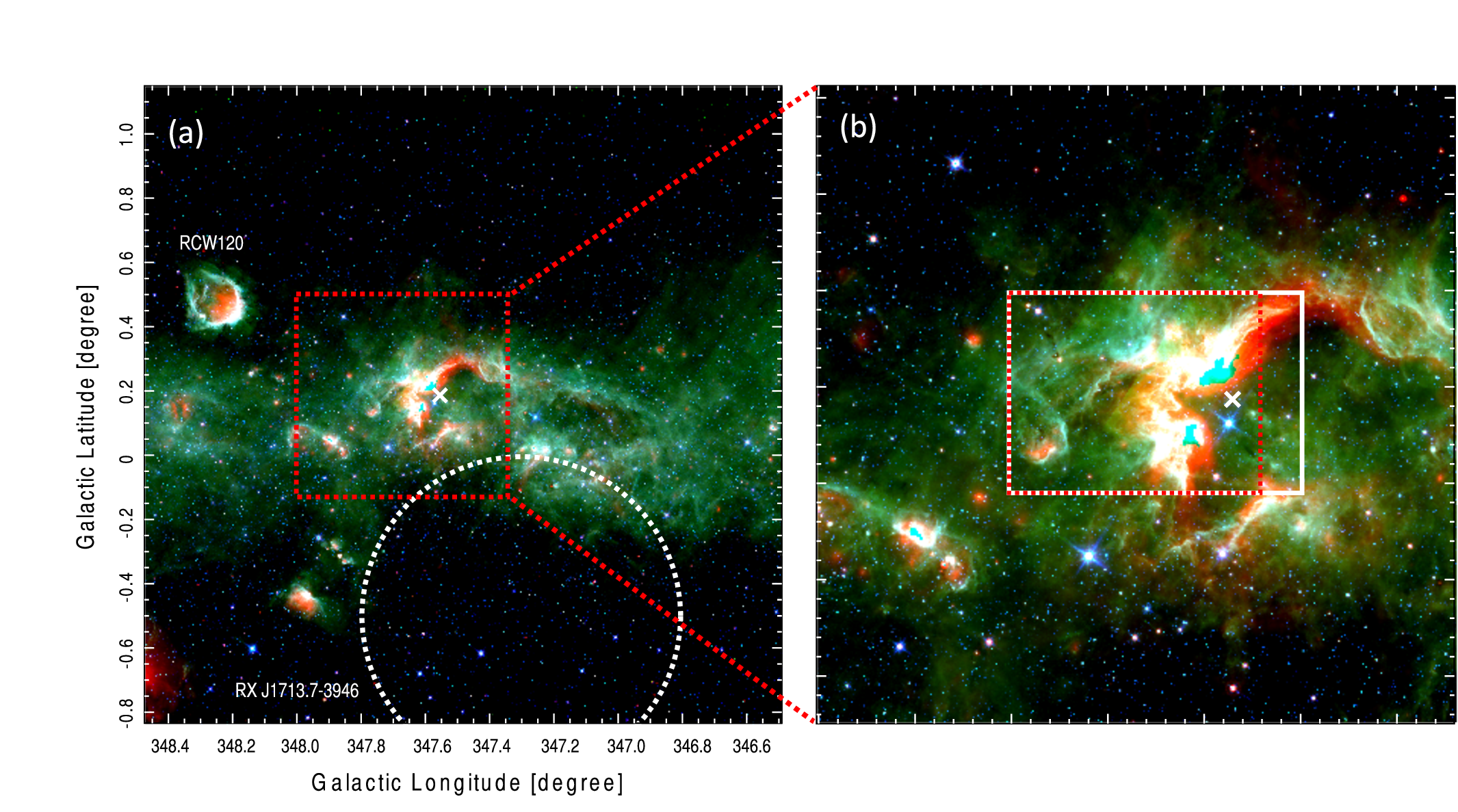}
  \end{center}
  \caption{(a) Composite color image of the {\it Spitzer}/IRAC 5.8\,$\mu$m (blue), 8\,$\mu$m (green), and MIPSGAL 24$\mu$m (red). The white cross represent the cluster position. (b) The close-up figure of (a). The red rectangle indicates the observed area obtained with Mopra and ASTE. The white rectangle indicates the same area as Figure\,\ref{fig:Mopra13ASTE_LB}. 
  }\label{fig:3colormap}
\end{figure}

High-mass stars are defined as a star having mass greater than 8\,$M_{\odot}$, which is a threshold to evolve to a supernova. High-mass stars are influential to the interstellar space by injecting a large amount of energy as ultraviolet radiation and stellar winds in addition to the large kinetic energy and the heavy elements released by a supernova in the end of stellar evolution. It is therefore crucial to understand the formation of high-mass stars in our efforts to better understand the galaxy evolution. Observations of high-mass star formation is however difficult because high-mass stars are rare in the Universe as compared with the low-mass stars including the sun. 
The formation of high-mass stars is not a simple scale up version of the low-mass star formation which is based on mass accretion through the circumstellar disk (\cite{2007ARA&A..45..481Z}). A key factor in high mass star formation is the mass accretion rate. \citet{1987ApJ...319..850W} showed that high-mass star formation requires large mass accretion rate. The typical mass accretion rate in low-mass star formation is 10$^{-5}$\,$M_{\odot}$\,yr$^{-1}$ or less, while formation of a 100\,$M_{\odot}$ star requires a high rate of 10$^{-3}$\,$M_{\odot}$\,yr$^{-1}$. Observations show that some stars, e.g., R136a1 (e.g., \cite{2010MNRAS.408..731C}) and HD 93129A (e.g., \cite{2011MNRAS.415.3354C}), have $\sim$100--300\,$M_{\odot}$. A massive dense core is a probable precursor of high-mass star formation, and it was suggested that such cores may be forming under an external trigger instead of spontaneous collapse; for instance, \citet{1998ASPC..148..150E} discussed three mechanisms, ``globule squeezing'', ``collect \& collapse'', and ``cloud--cloud collision'' as possible ways to form high-mass stars, whereas there was no dedicated work along this direction in the literature. In spite of these works as well as many others it is not yet settled how high-mass stars form (e.g., \cite{2007ARA&A..45..481Z}, \cite{2014prpl.conf..149T}) . 
In external galaxies Super Star Clusters (SSCs) (\cite{1985AJ.....90.1163A}) were known as an outstanding massive stellar system which may be similar to old globular clusters. In the Milky Way Super Star Clusters have $10^{4}$\,$M_{\odot}$ and harbor tens of O stars in a pc volume (see for a review\cite{2010ARA&A..48..431P}), corresponding to stellar density of $10^{4}$\,pc$^{-3}$ (\cite{2005IAUS..227..413J}). Many O stars in a SSC strongly ionize the surrounding molecular gas and it is usually difficult to identify the parent cloud of a SSC. Nonetheless, \citet{2009ApJ...696L.115F} and \citet{2010ApJ...709..975O} discovered two parent molecular clouds in collision in Westerlund 2, a SSC with an age of 2\,Myr, by large-scale CO observations with NANTEN2. Subsequently, \cite{2014ApJ...780...36F};\yearcite{2016ApJ...820...26F} discovered two colliding parent clouds in SSCs NGC3630 and RCW38, both of which are younger than 2\,Myr. These works raised a possibility that formation of a SSC may be triggered by strong gas compression in cloud--cloud collision, and a further search for parent clouds in SSCs was strongly desired. Possible candidates for the search are young SSCs including Arches (\cite{1999ApJ...525..750F}), Quintuplet (\cite{1999ApJ...525..750F}), Trumpler 14 (\cite{2007A&A...476..199A}), and [DBS2003]179 (\cite{2008A&A...488..151B},\yearcite{2012A&A...546A.110B}). 
[DBS2003]179 is a super star cluster discovered by 2MASS (\cite{2003A&A...400..533D}) at $(l,b)=(\timeform{347.58D}, \timeform{0.19D})$ and is associated with an H{\sc ii} region cataloged by Caswell and Haynes 1987. An H{\sc ii} region is identified at $(l,b)(ep=J2000) = (\timeform{347.600D} , \timeform{+0.211D})$ by \cite{1987A&A...171..261C}. \cite{1970A&A.....6..364W} derived kinematic distance of 7.9$\pm$1.0\,kpc and 11.6$\pm$1.0\,kpc based of the observed H109α velocity -96.2$\pm$0.7\,km\,s$^{-1}$. Distance is estimated to be 7.9\,kpc (\cite{2012A&A...546A.110B}), 6.4\,kpc, (\cite{2011AJ....142...40M}) and 9.0\,kpc (\cite{2012MNRAS.419.1860D}). Figures\,\ref{fig:3colormap}(a)(b) show a three-color image toward [DBS2003]179 obtained with {\it Spitzer} space telescope at 5.8\,$\mu$m, 8\,$\mu$m, and 24\,$\mu$m. The bright features indicate association with the cluster. A gamma ray SNR RX J1713.7-3936 (Pfeffermann \& Aschenbach 1996) and a {\it Spitzer} Bubble RCW120 (\cite{1960MNRAS.121..103R}) are located close to the cluster in the sky. \cite{2008A&A...488..151B};\yearcite{2012A&A...546A.110B} estimated the cluster mass to be $2.5 \times 10^{4}$\,$M_{\odot}$ and a size of 0.4\,pc. The age of the cluster is estimated to be 2--5\,Myrs (\cite{2012A&A...546A.110B}). \citet{2011AJ....142...40M} used color selection by {\it Spitzer}/GLIMPSE and 2MASS database and X ray point source catalogs of Chandra X--ray Observatory (\cite{2010ApJS..189...37E}) XMM--Newton (\cite{2004yCat..73510031H}) more than 10 O stars having 20\,$M_{\odot}$, and 3 WR stars WR stars WN8-9 MDM32 etc. having more than 85 Mo are included (\cite{2011AJ....142...40M}). 
The aim of the present paper is to identify the molecular clouds associated with [DBS2003]179 and test if cloud--cloud collision is working to form the cluster as found in the other super star clusters in the Milky Way. The paper is organized as follows; Section 2 gives description of the observations with NANTEN2, ASTE and the Mopra 22m telescope. Section 3 gives the observational results at multi wavelengths and the physical parameters of the molecular clouds as well as the outcomes derived from a LVG analysis. Section 4 we discuss the association of the molecular clouds and examine the formation process of [DBS2003]179 from a view point of cloud--cloud collision. Finally, we compare DBS179 with the other SSCs and discuss cluster formation and cloud--cloud collisions. Section 5 gives conclusions of the present study.

\section{Dataset}

\subsection{NANTEN2}\label{sec:dat_n}
The molecular data of the $^{12}$CO ($J$=1--0) and $^{13}$CO ($J$=1--0) transitions were obtained with the 4m mm and sub-mm telescope NANTEN2 located in Atacama at 4859m above sea level. observation were made from 2012 May to 2012 December. These observations were made in the On-The-Fly (OTF)  mode. The grid spacings were 60" and the Half Power Beam Width was 180$''$. In the present paper we used the data within $1\fdg7\times 1\fdg6$ of [DBS2003]179. We employed a 100\,GHz 4\,K SIS mixer receiver which has Double-Side-Band system temperature (DSB) of $\sim$250K toward the zenith. Spectroscopy was made with a digital spectrometer of 1\,GHz band width having 16384\,channels the frequency width of each channel is 0.06\,MHz. The total velocity width was 2600\,km\,s$^{-1}$ with 0.17\,km\,s$^{-1}$ smoothed to velocity resolution of 0.11\,km\,s$^{-1}$. 
Pointing was monitored by observing the sun and IRC+10216 ($\alpha_{\rm J2000},\delta_{\rm J2000})  = (\timeform{9h47m57s.406}, \timeform{13D16'43''.56})$ everyday and the error was measured to be less than 15$''$ in rms. The intensity scale was established by combining the room temperature load and the sky emission, and the absolute intensity scale was established by observing IRAS 16293--2422 $(\alpha_{\rm J2000},\delta_{\rm J2000})  = (\timeform{16h32m23s.3}, \timeform{-24D28'39''.20})$ in LDN 1689 and the nearby Perseus molecular cloud $(\alpha_{\rm J2000},\delta_{\rm J2000})  = (\timeform{3h29m19s.0}, \timeform{31D24'49''.0})$. The main beam efficiency of NANTEN2 was measured to be 0.53, and the typical noise fluctuations were 0.8\,K and 0.5\,K for 0.11\,km\,s$^{-1}$ velocity resolution for $^{12}$CO ($J$=1--0) and $^{13}$CO ($J$=1--0), respectively. The observation with NANTEN2 were compared with the smoothed data taken by Five College Radio Astronomy Observatory (FCRAO). 

\subsection{Mopra 22m telescope}\label{sec:dat_m}
We used the high resolution data of $^{12}$CO ($J$=1--0), $^{13}$CO ($J$=1--0) and C$^{18}$O ($J$=1--0) emissions obtained with the Mopra 22m telescope in NSW of Australia. The observations were made from 2012 May to October in the On-The-Fly (OTF) mode. Grid spacing was taken to be 15" for a Half Power Beam Width (HPBW) of 33$''$(\cite{2005PASA...22...62L}). Observations were made several times for a tile of $4' \times 4'$ in total a field of $16'\times 13'$ toward the cluster was used of [DBS2003]179. The receiver was a 3\,mm MMIC HEMT receiver covering a frequency range 76--117\,GHz. The system noise temperature was $\sim$600\,K at the $^{12}$CO ($J$=1--0) frequency and $\sim$250\,K in the $^{13}$CO ($J$=1--0) and C$^{18}$O ($J$=1--0) frequency band.  

The spectroscopy was made with a digital spectrometer MOPS which has both wide-band mode and zoom mode setups. The present observations were made in the zoom mode for 4 sub-bands and a 8.3\,GHz band in total was covered simultaneously.  Each sub-band observes 2.3\,GHz window for the both polarizations with 8096\,frequency\,channels. The zoom mode covers 137.5\,MHz for 16\,frequency\,bands with dual polarizations with 4096\,channels, where each channel has a 0.03\,MHz width. This corresponds to 376\,km\,s$^{-1}$ with a velocity channel of 0.09\,km\,s$^{-1}$. The effective velocity resolution is 0.11\,km\,s$^{-1}$ after smoothing. Pointing accuracy was achieved by observing SiO masers. They are $ (\alpha_{\rm J2000},\delta_{\rm J2000})  = (\timeform{18h08m04s.048}, \timeform{-22D13'26''.63})$ and a red giant AH Sco $(\alpha_{\rm J2000},\delta_{\rm J2000})  = (\timeform{17h11m17s.021}, \timeform{-32D19'30''.71})$ observed every 1-2 hr by the five-point method. The pointing was accurate within 10", the absolute intensity scale was established by observing 

M17SW $(\alpha_{\rm J2000},\delta_{\rm J2000})  = (\timeform{18h320m23s.1}, \timeform{-16D11'43''})$ M17SW $^{12}$CO ($J$=1--0) average brightness temperature of M17SWbeam efficiency was estimated to be  0.49 by using the observed value by \citet{2005PASA...22...62L}. The typical rms noise fluctuations were 1.4K, 0.3K, and 0.4K for $^{12}$CO ($J$=1--0), $^{13}$CO ($J$=1--0), and C$^{18}$O ($J$=1--0) , respectively in a velocity channel of 0.11\,km\,s$^{-1}$.

\subsection{ASTE observations}\label{sec:dat_a}
The $^{12}$CO ($J$=3--2) data used in the present paper were taken with ASTE 10 m telescope in Atacama, Chile (\cite{2004SPIE.5489..763E}, \cite{2004dimg.conf..349K}). Observations were made in June 2014 by using the $^{12}$CO ($J$=3--2) emission in the On-The-Fly (OTF) mode (e.g, \cite{2007A&A...474..679M}; \cite{2008PASJ...60..445S}) with a grid interval of 10$''$ with a main beam of 22$''$ in half power beam width (HPBW).  Observations were made for several tiles of a $4'\times 4'$ field in RA and Dec toward [DBS2003]179, and the final coverage was $16'\times 13'$ . The front wand was CATS345 2SB SIS receiver (\cite{2008SPIE.7012E..08E}, \cite{2008stt..conf..281I}) which covers a frequency range of 324--372\,GHz. The system noise temperature was 180--250K in DSB in the $^{12}$CO ($J$=3--2) frequency. Spectroscopy was made with a digital spectrometer MAC of the XF type having the 512\,MHz mode and the 128\,MHz mode (\cite{2000SPIE.4015...86S}). The present observations used the 128\,MHz mode, where the signal is fed into 1024\,channels of 0.125\,MHz channel width. These correspond to velocity coverage of 111\,km\,s$^{-1}$ and a velocity channel of 0.11\,km\,s$^{-1}$. Pointing accuracy was monitored by observing a post-AGB Star IRAS 16594-4656 $(\alpha_{\rm J2000},\delta_{\rm J2000})  = (\timeform{17h03m10s.027}, \timeform{-47D00'27''.68})$ every 1.5 hr by the five point method. Error was found to be within 2$''$. The absolute intensity scale referred to W28 $(\alpha_{\rm B1950},\delta_{\rm B1950})  = (\timeform{17h57m26s.8}, \timeform{-24D03'54''.0})$ whose $^{12}$CO ($J$=3--2) intensity was compared with the average brightness observed by \cite{1994ApJS...95..503W}) and we estimated a beam efficacy of 0.60. The final RMS noise fluctuations were 0.4\,K with a velocity resolution of 0.11\,km\,s$^{-1}$.

\clearpage

\section{Results}\label{sec:res}

\subsection{CO distributions}\label{sec:res_CO}
\begin{figure}
  \begin{center}
  \includegraphics[width=\hsize]{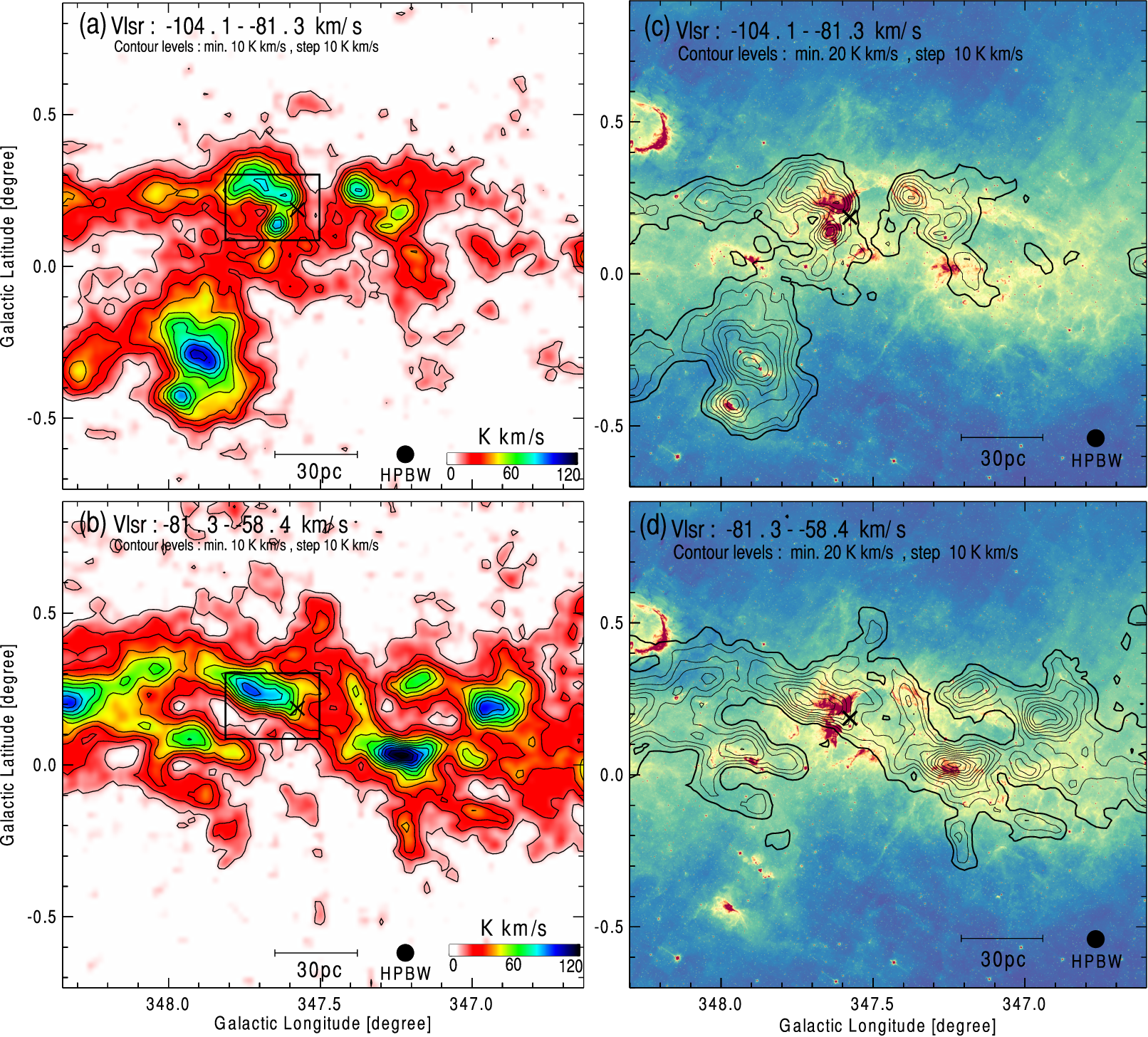}
  \end{center}
  \caption{(a) (b) The $^{12}$CO ($J$=1--0) integrated intensity map obtained with NANTEN2 between the velocities of $-$104.1\,km\,s$^{-1}$ and $-$81.3\,km\,s$^{-1}$ for (a) and between the velocities of $-$81.3\,km\,s$^{-1}$ and $-$60.8\,km\,s$^{-1}$ for (b), respectively. The contours are plotted at every 10\,K\,km\,s$^{-1}$ from 10\,K\,km\,s$^{-1}$. The cross represent the cluster position. The black rectangle indicates the area of Figure\,\ref{fig:Mopra13ASTE_LB}. (c) (d) The {\it Spitzer}/IRAC 8\,$\mu$m image. The contours are plotted at every 10\,K\,km\,s$^{-1}$ from 20\,K\,km\,s$^{-1}$.
  }\label{fig:NANTEN2}
\end{figure}

Figures\,\ref{fig:NANTEN2}(a)(b) show velocity channel distributions of a large area toward DBS2003[179] in the $^{12}$CO ($J$=1--0) emission obtained with NANTEN2. The region is close to the Galactic center and we see many CO features, where we focus on the two CO clouds which show intense emission close to the cluster at two velocity ranges of -90\,km\,s$^{-1}$ and -70\,km\,s$^{-1}$. Figures\,\ref{fig:NANTEN2}(c)(d) shows the two CO components and their superposition with the {\it Spitzer} infrared image at 8\,$\mu$m. We find three CO peaks lie within 30\,pc of the cluster indicating candidates for physically associated molecular gas to the cluster. 

\begin{figure}[htbp]
  \begin{center}
  \includegraphics[width=14cm]{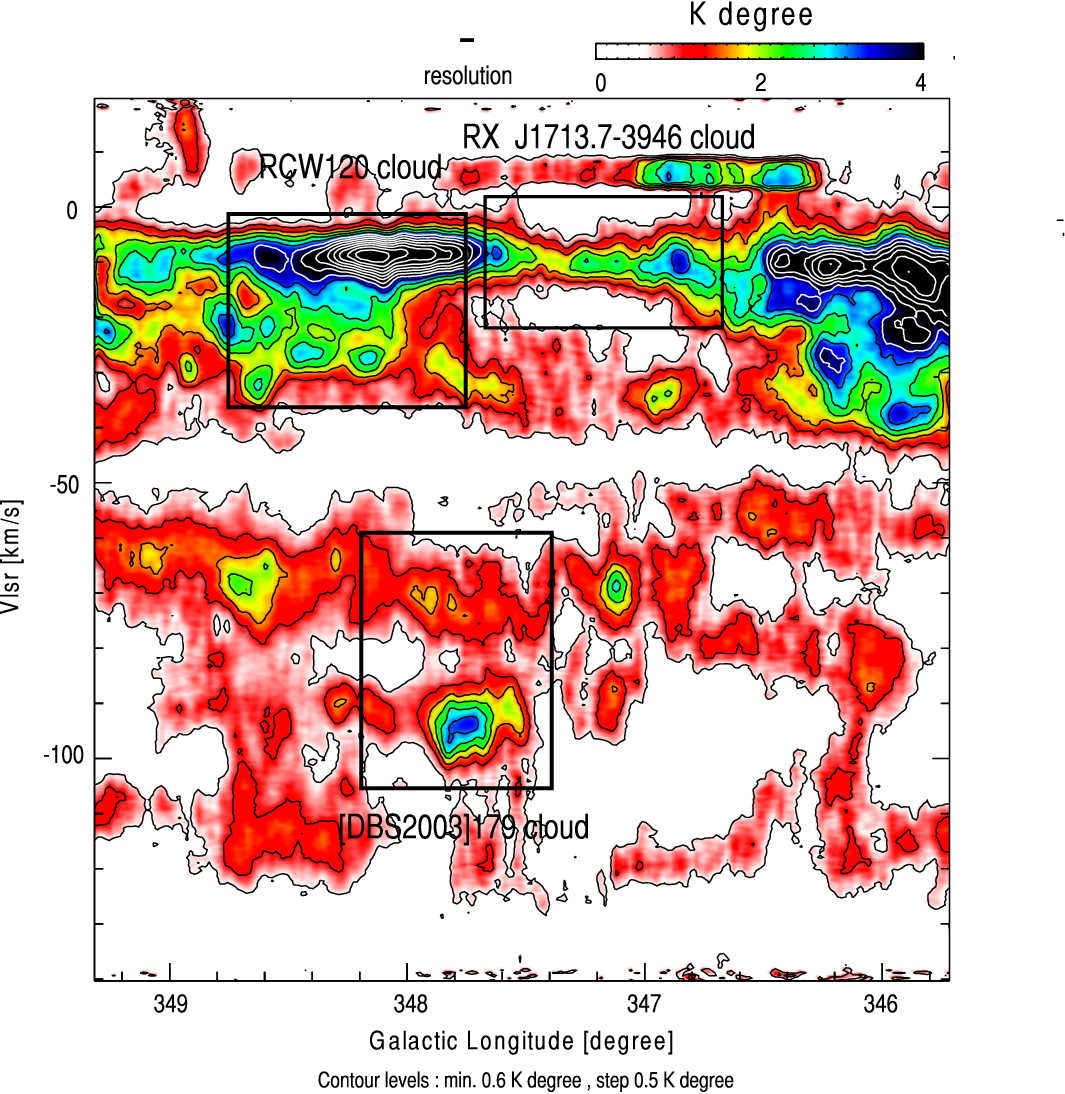}
  \end{center}
  \caption{The $^{12}$CO ($J$=1--0) $l$--$v$ diagram obtained with NANTEN2 between the $b$ of $-$1.0 and $+$1.0 degrees. The black rectangles indicate the molecular clouds associating with RCW120 (\cite{2015ApJ...806....7T}) and RX J1713.7-3936 (\cite{2010ApJ...724...59S, 2013ApJ...778...59S}), respectively. 
  }\label{fig:NANTEN12CO-wideLV}
\end{figure}

Figure\,\ref{fig:NANTEN12CO-wideLV} shows a longitude-velocity diagram and we indicate the two velocity clouds by a box. The direction is by chance close to the other two known objects, a {\it Spitzer} bubble RCW120 and a gamma ray SNR RXJ1713. The CO emission which are likely associated with [DBS2003]179 shows two peaks at -94\,km\,s$^{-1}$ and at -72\,km\,s$^{-1}$ over $\sim$0.4\,degrees in l, and becomes weak in a velocity range -86.1\,km\,s$^{-1}$ - -78.9\,km\,s$^{-1}$. The two velocity components show a bridge feature connecting them in velocity.

\begin{figure}[htbp]
  \begin{center}
  \includegraphics[width=14cm]{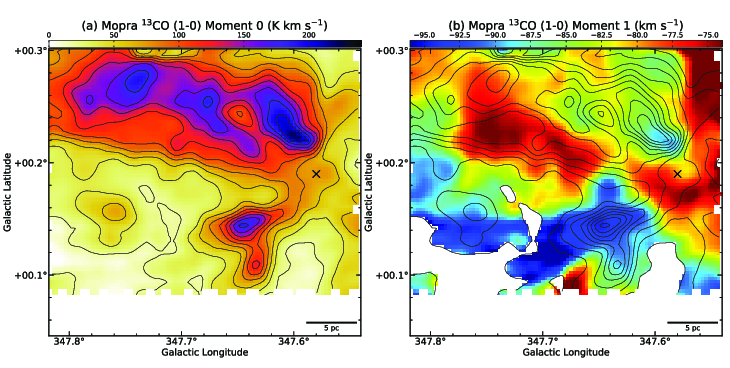}
  \end{center}
  \caption{(a) The 0th moment of the $^{12}$CO ($J$=1--0) emission obtained with Mopra between the velocities of $-$104.1\,km\,s$^{-1}$ and $-$60.8\,km\,s$^{-1}$. The contours are plotted at every 20\,K\,km\,s$^{-1}$ from 20\,K\,km\,s$^{-1}$. (b) The 1st moment of the $^{12}$CO ($J$=1--0) emission obtained with Mopra. The velocity range and contour levels are the same as (a). }\label{fig:moms}
\end{figure}

\begin{figure}[htbp]
  \begin{center}
  \includegraphics[width=14cm]{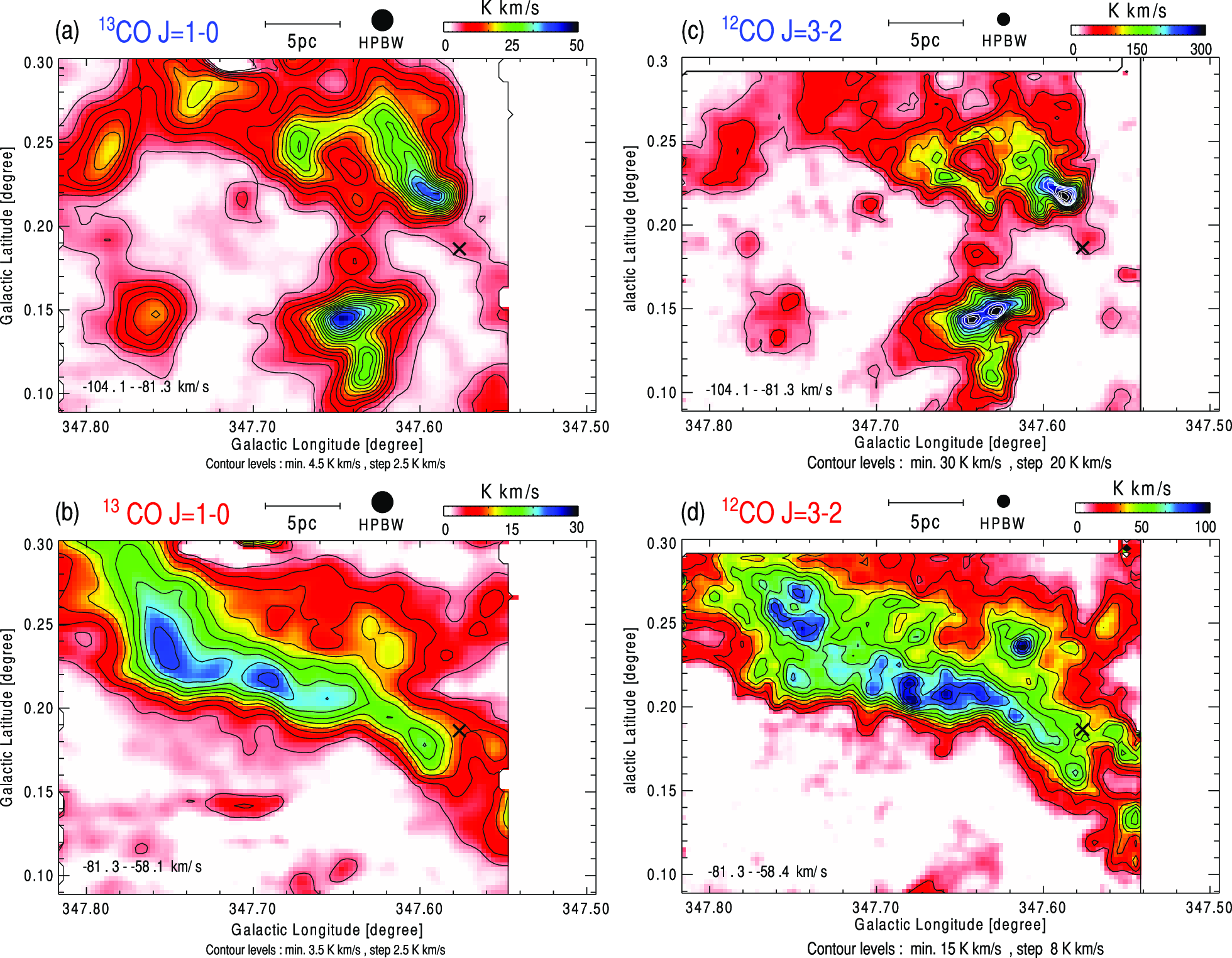}
  \end{center}
  \caption{(a) The $^{13}$CO ($J$=1--0) integrated intensity map obtained with Mopra between the velocities of $-$104.1\,km\,s$^{-1}$ and $-$81.3\,km\,s$^{-1}$. The contours are plotted at every 2.5\,K\,km\,s$^{-1}$ from 4.5\,K\,km\,s$^{-1}$. (b) The same as (a), but between the velocities of $-$81.3\,km\,s$^{-1}$ and $-$60.8\,km\,s$^{-1}$. The contours are plotted at every 2.5\,K\,km\,s$^{-1}$ from 3.5\,K\,km\,s$^{-1}$. (c) The $^{12}$CO ($J$=3--2) integrated intensity map obtained with ASTE between the velocities of $-$104.1\,km\,s$^{-1}$ and $-$81.3\,km\,s$^{-1}$. The contours are plotted at every 30\,K\,km\,s$^{-1}$ from 20\,K\,km\,s$^{-1}$. (d) The same as (c), but between the velocities of $-$81.3\,km\,s$^{-1}$ and $-$60.8\,km\,s$^{-1}$. The contours are plotted at every 8\,K\,km\,s$^{-1}$ from 15\,K\,km\,s$^{-1}$. 
  }\label{fig:Mopra13ASTE_LB}
\end{figure}

Figures\,\ref{fig:moms}(a)(b) show the moment 0 and moment 1 map of the $^{12}$CO ($J$=1--0) emission obtained with Mopra telescope, respectively, in a velocity range from -104.1 to -58.4\,km\,s$^{-1}$. 
The moment 1 map obtained with Mopra telescope also indicates that the CO emission in this velocity range can be separated into two velocity components. 
Figure\,\ref{fig:Mopra13ASTE_LB} shows the distribution of the $^{13}$CO (J=1--0) and $^{12}$CO ($J$=3--2) emission for the two velocity components obtained with Mopra and ASTE, respectively. 
In particular, the J=3--2 emission reveals details at a high resolution of $\sim22$\,arcsec.

\begin{figure}[htbp]
  \begin{center}
  \includegraphics[width=10cm]{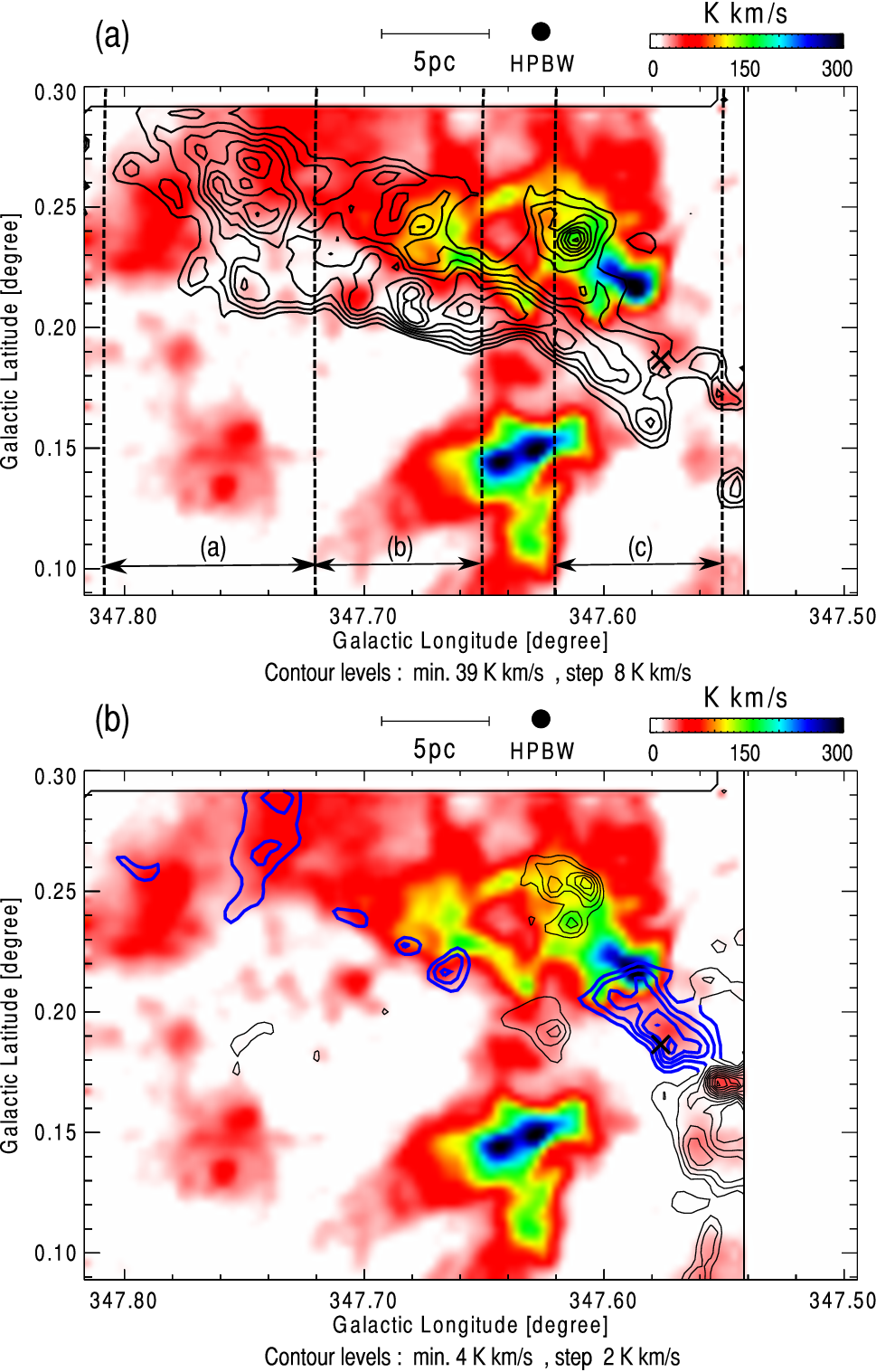}
  \end{center}
  \caption{(a) The $^{12}$CO ($J$=3--2) integrated intensity map. The color scale and the contours show the distribution of the blue-shifted cloud and the red-shifted cloud, respectively. The contours are plotted at every 8\,K\,km\,s$^{-1}$ from 39\,K\,km\,s$^{-1}$. (b) The black and blue contours show the $^{12}$CO ($J$=3--2) integrated intensity between the velocities of $-$83\,km\,s$^{-1}$ and $-$81\,km\,s$^{-1}$, and are plotted at every 2\,K\,km\,s$^{-1}$ from 4\,K\,km\,s$^{-1}$. The blue contours indicate bridge components.
  }\label{fig:ASTEBridge}
\end{figure}

\begin{figure}[htbp]
  \begin{center}
  \includegraphics[width=9cm]{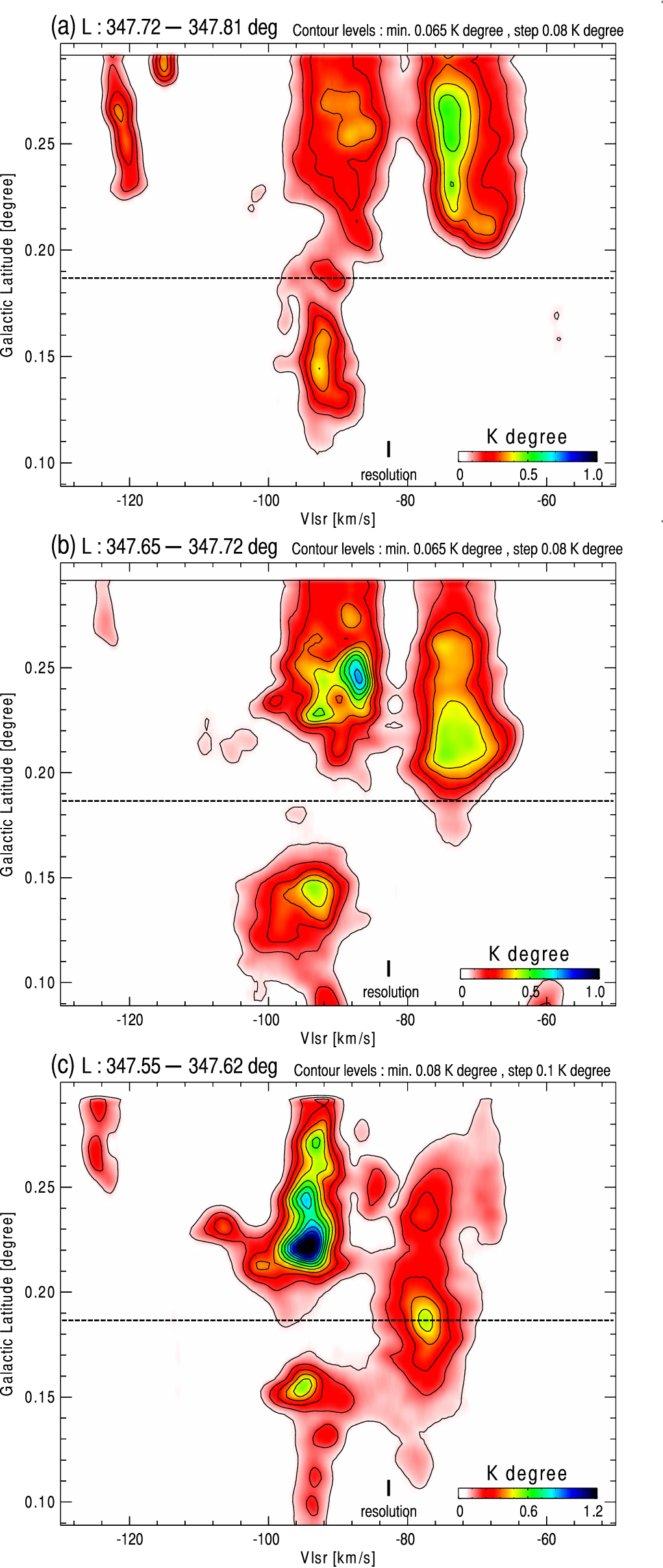}
  \end{center}
  \caption{The $^{12}$CO ($J$=3--2) $v$--$b$ diagram obtained with ASTE between the $l$ of (a) $347\fdg81$ and $347\fdg72$, (b) $347\fdg72$ and $347\fdg65$, and (c) $347\fdg62$ and $347\fdg55$. The horizontal dashed-line indicates the position of the star cluster.
  }\label{fig:ASTE_BVdiagram}
\end{figure}

Figure\,\ref{fig:ASTEBridge}(a) shows an overlay of the two velocity CO components the blue-shifted cloud and the red-shifted cloud, and Figure\,\ref{fig:ASTEBridge}(b) an overlay with the bridge features in a velocity range of -83 \,km\,s$^{-1}$ -- -81 \,km\,s$^{-1}$. Figure\,\ref{fig:ASTEBridge}(a) indicates that the two components show complementary distribution with each other in the sense that the red-shifted cloud is distributed between the two elongated features of the blue-shifted clouds along the Galactic plane over 0.2\,degrees. The bridge features are distributed within 0.05\,degrees of the cluster and toward the interface between the two components at $l\sim$347.65--347.75. Figure\,\ref{fig:ASTE_BVdiagram} shows velocity--latitude diagrams which include the bridge features at three longitude ranges, $347\fdg72--347\fdg81$, $347\fdg65--347\fdg72$ and $347\fdg55--347\fdg62$, and shows that the bridge features are connecting the blue- and red-shifted clouds. 

\begin{figure}[htbp]
  \begin{center}
  \includegraphics[width=14cm]{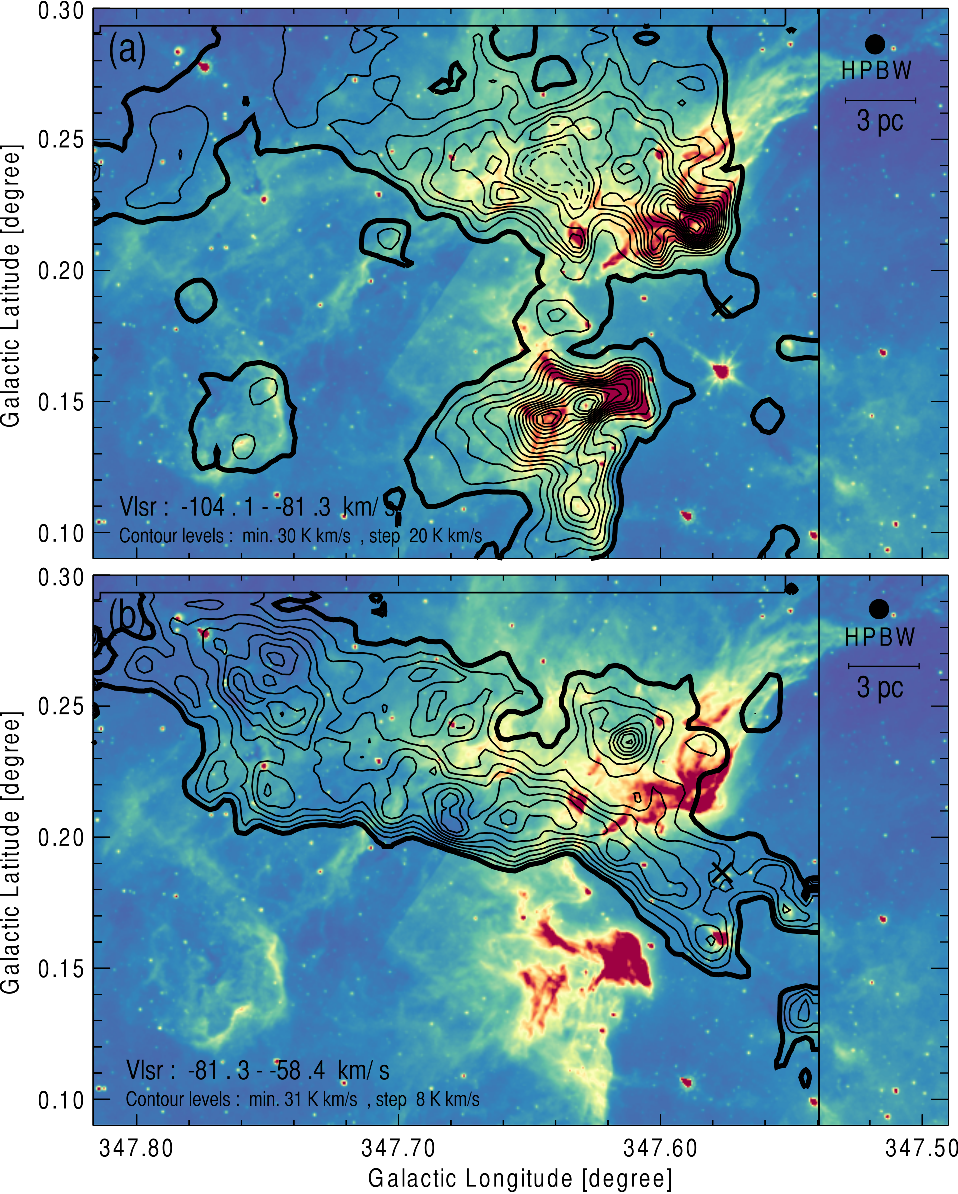}
  \end{center}
  \caption{(a) The {\it Spitzer}/IRAC 8\,$\mu$m image. The contours show the $^{12}$CO ($J$=3--2) integrated intensity of the blue-shifted cloud, and are plotted at every 20\,K\,km\,s$^{-1}$ from 30\,K\,km\,s$^{-1}$. The dashed contour indicates an intensity decreasing. (b) The same as (a), but for the red-shifted cloud. The contours are plotted at every 8\,K\,km\,s$^{-1}$ from 31\,K\,km\,s$^{-1}$. 
  }\label{fig:ASTESpitzer8}
\end{figure}

\begin{figure}[htbp]
  \begin{center}
  \includegraphics[width=14cm]{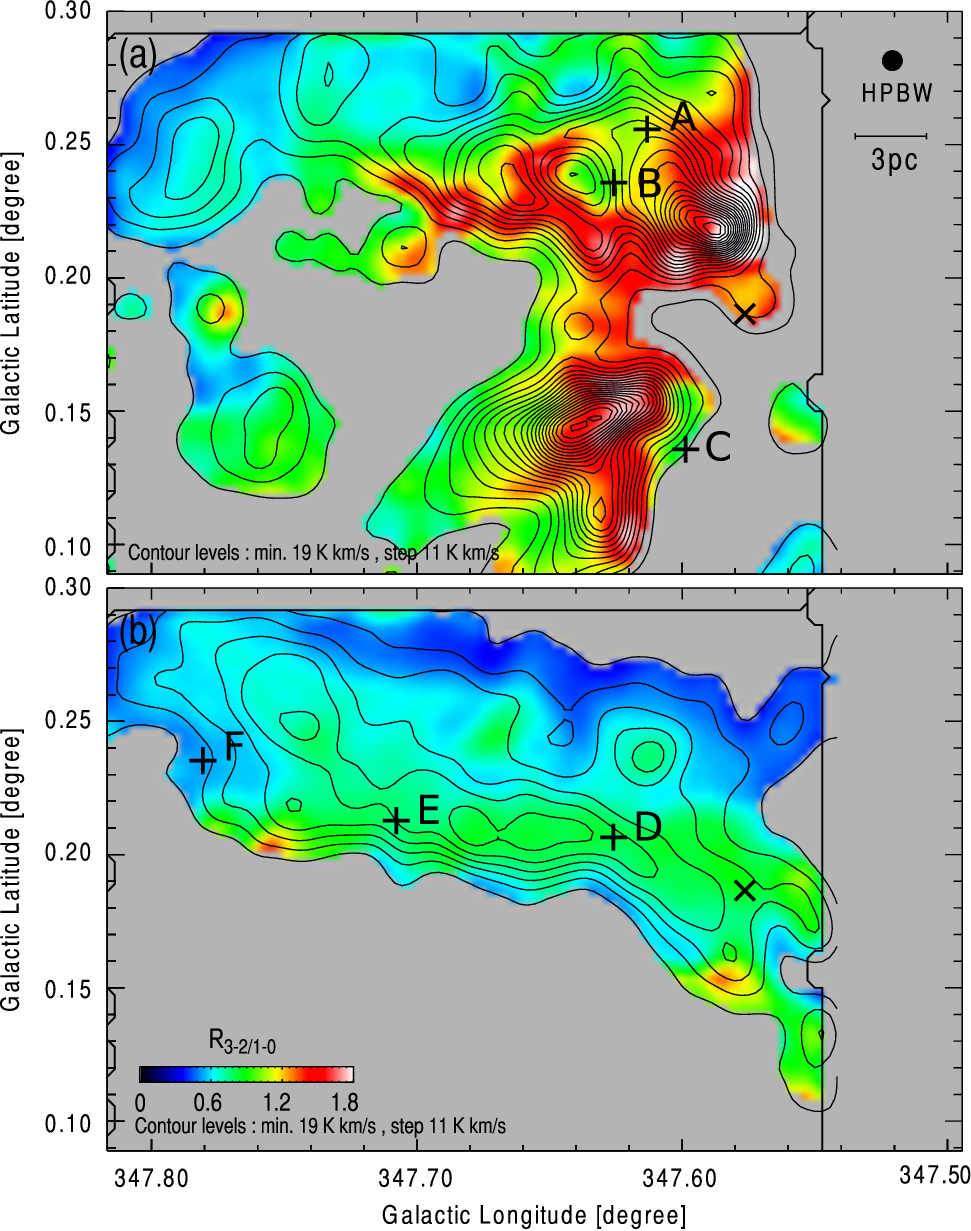}
  \end{center}
  \caption{(a) (b) The $^{12}$CO ($J$=3--2)/$^{12}$CO ($J$=1--0) integrated intensity ratio map for blue-shifted cloud and red-shifted cloud, respectively. The contours are plotted at every 11\,K\,km\,s$^{-1}$ ($\sim 3\sigma$) from 19\,K\,km\,s$^{-1}$ ($\sim 5\sigma$). The crosses A--F represent the positions where we applied a LVG analysis in Section \ref{sec:res_LVG}. 
  }\label{fig:Ratio3-2.1-0}
\end{figure}

Figure\,\ref{fig:ASTESpitzer8} shows a comparison of the two velocity components with the {\it Spitzer} 8\,$\mu$m image. The red-shifted cloud shows a strong correlation with the 8\,$\mu$m image indicates that the cloud is close to the cluster and is illuminated at a small distance within 10\,pc degrees. Conversely, the blue-shifted cloud shows no strong illumination by the cluster suggesting that the cloud is located at a distance more than 10\,pc, except for a small peak at $(l,b=347\fdg58,0\fdg17)$ showing a high ratio of 0.8, which is possibly located a distance around 10\,pc of the cluster. Another piece of evidence for association of the clouds with the cluster is obtained by the ratio distribution of the CO 3--2/1--0 intensity in Figure\,\ref{fig:Ratio3-2.1-0}. The red-shifted cloud shows high ratios of 0.8--2.0, conversely, the blue shifted cloud shows a ratio of 0.8--1.3 which is still considerably higher than 0.6. The typical ratio in molecular clouds without extra heating shows a ratio of $\sim$0.4 (e.g., \cite{2007PASJ...59...15O}), and a higher ratio above 0.6 indicates extra-heating by the cluster.

\subsection{LVG analysis}\label{sec:res_LVG}
\begin{figure}[htbp]
  \begin{center}
  \includegraphics[width=14cm]{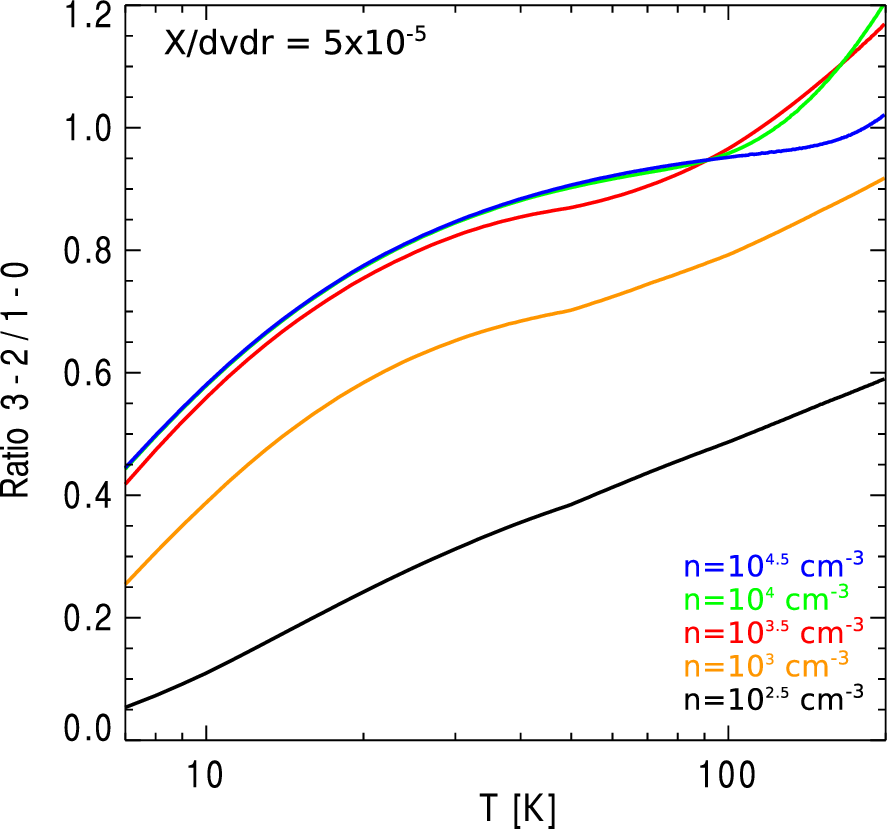}
  \end{center}
  \caption{The curves of the $^{12}$CO ($J$=3--2)/$^{12}$CO ($J$=1--0) integrated intensity ratio as a function of $T_{\rm k}$ and $n({\rm H_2})$, estimated using the LVG calculations. $\it X$$_{CO}$$\it / (dv/dr)$ is assumed as $5 \times 10^{-5}$.  
  }\label{fig:Ratio3-2.1-0_1LVG}
\end{figure}

The CO spectra are collisionally exited by H$_2$ molecules as a main collision partner. Figure\,\ref{fig:Ratio3-2.1-0_1LVG} shows the brightness temperature ratio of the $J$=3--2/1--0 transition of $^{12}$CO. Density of the present two clouds is likely above $10^{3.5}$\,cm$^{-3}$, and the ratio becomes above 0.8 for kinetic temperature higher than 20\,K. The usual temperature of molecular clouds is 10\,K under cosmic ray heating alone, and 20\,K requires extra heating usually by high mass stars. For the ratio of the red component $R_{3-2/1-0}=0.8$, temperature exceeds 10\,K, which is the typical cloud temperature without local extra heating, suggests that heating by the cluster is working.

We applied the Large Velocity Gradient analysis (\cite{1974ApJ...189..441G}) in order to estimate density and temperature of the molecular clouds. We adopt the parameters in the analysis, the $^{12}$CO / $^{13}$ ratio 53 (\cite{1994ARA&A..32..191W}) and the $^{12}$CO abundance $X_{\rm CO} = 10^{-4}$ (\cite{1982ApJ...262..590F} and \cite{1984ApJS...56..231L}). We assumed a velocity gradient along the line of sight of $dv=dr 5-10 $\,km\,s$^{-1}$/pc by taking an average among the clumps in the clouds. The error in intensity is taken to be 10 \% in $^{12}$CO ($J$= 1--0) and $^{12}$CO ($J$=3--2), while the error is negligibly small in $^{12}$CO ($J$= 1--0) $^{13}$CO ($J$= 1--0). In order to avoid possible effects of self-absorption in $^{12}$CO we inspected carefully the line profiles and excluded peak positions where self-absorption may be present.

\begin{figure}[htbp]
  \begin{center}
  \includegraphics[width=14cm]{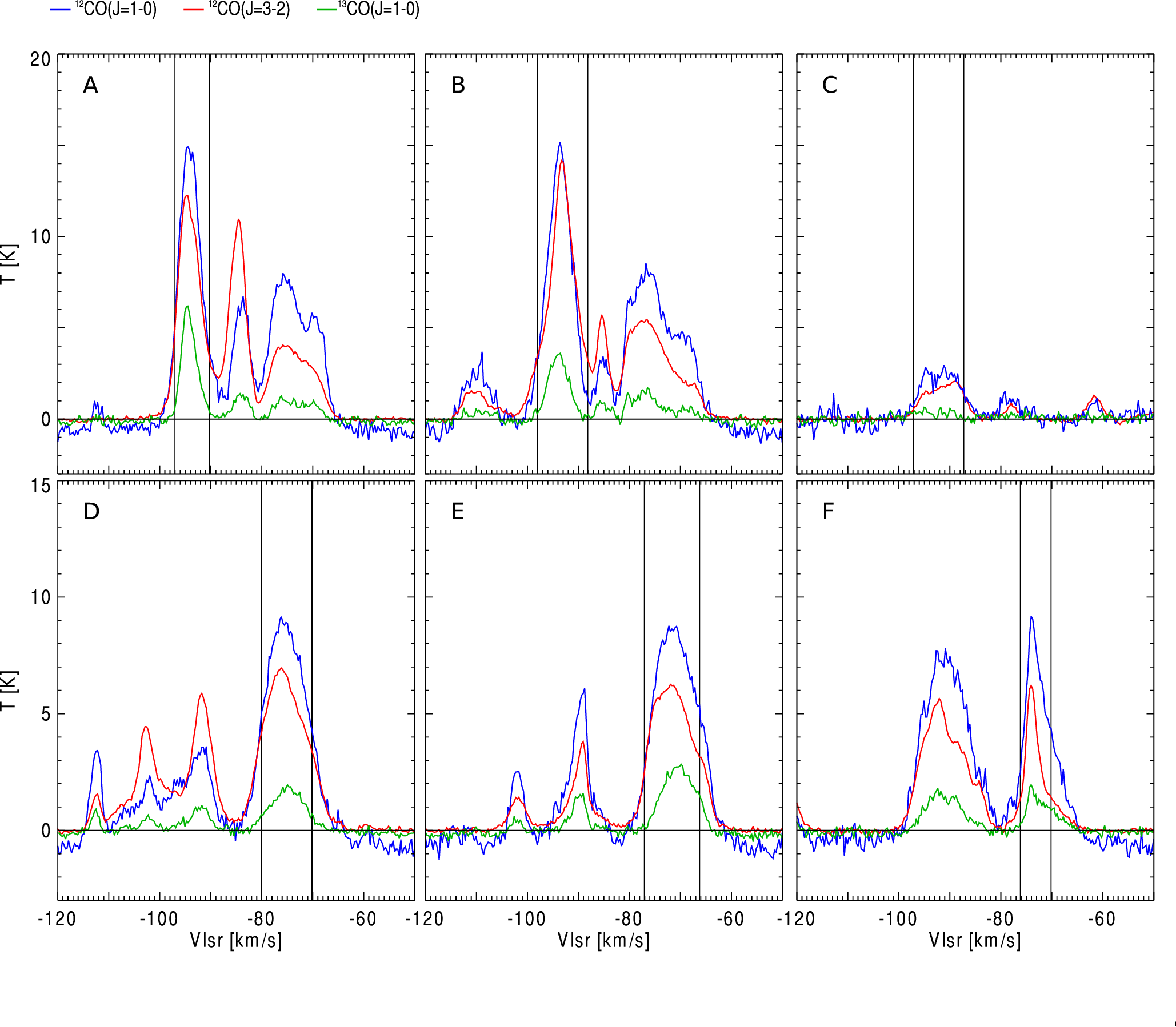}
  \end{center}
  \caption{The CO emission line profiles at the position A--C (the blue-shifted cloud) and D--F (the blue-shifted cloud) shown in Figure\,{fig:Ratio3-2.1-0}. The blue, red, and black lines indicate Mopra $^{12}$CO ($J$=1--0), ASTE $^{12}$CO ($J$=3--2), and Mopra $^{13}$CO ($J$=1--0) intensity, respectively. The black vertical lines indicate the velocity range applied for the LVG analyses.
  }\label{fig:LVG_LineProfile}
\end{figure}

\begin{figure}[htbp]
  \begin{center}
  \includegraphics[width=14cm]{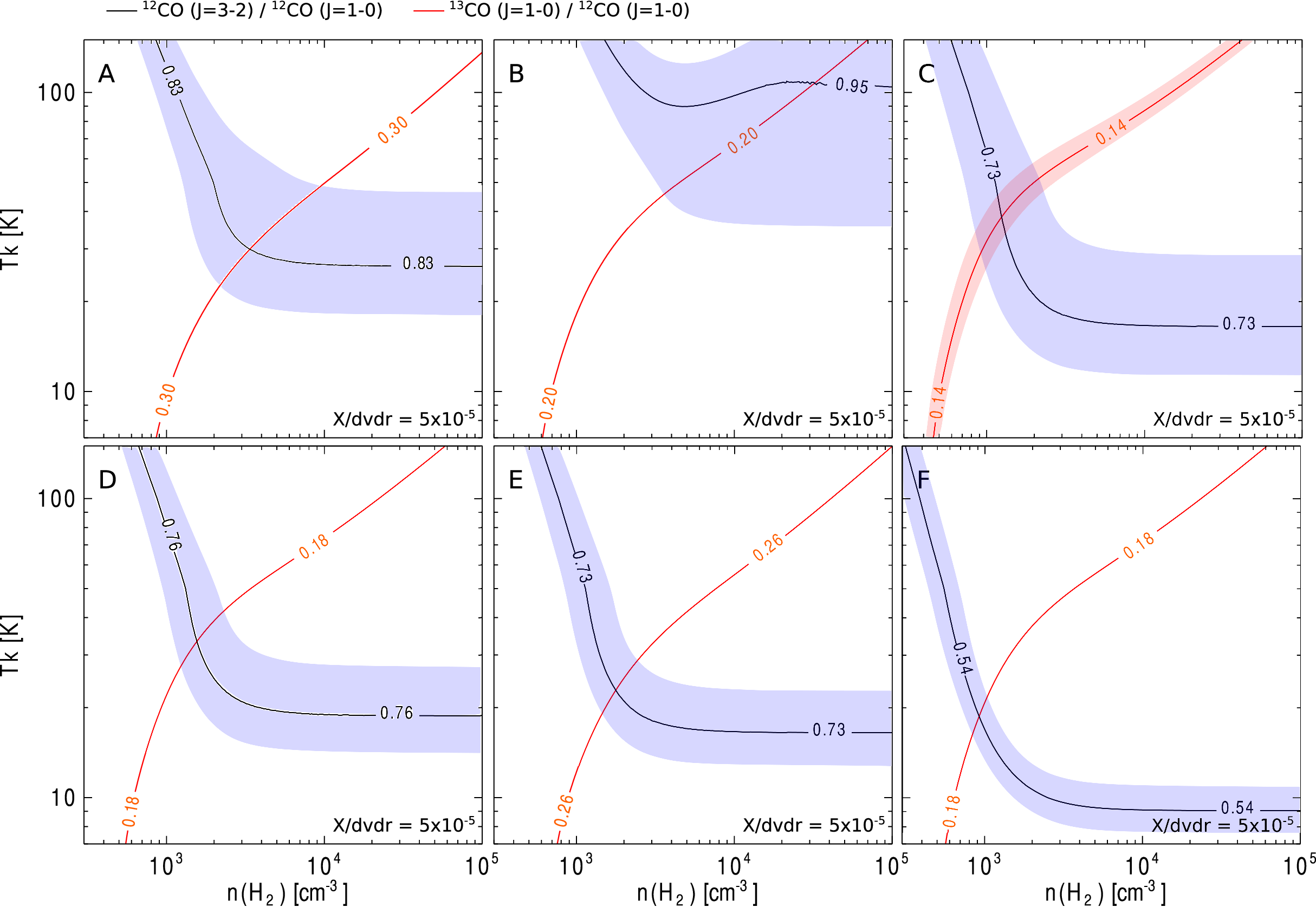}
  \end{center}
  \caption{The results of the LVG analyses at the position A--C (the blue-shifted cloud) and D--F (the blue-shifted cloud) shown in Figure\,{fig:Ratio3-2.1-0}. The black lines and the blue area show the $^{12}$CO ($J$=3--2)/$^{12}$CO ($J$=1--0) integrated intensity ratio and their error range, respectively. The red lines and the red area show the $^{13}$CO ($J$=1--0)/$^{12}$CO ($J$=1--0) integrated intensity ratio and their error range, respectively. $\it X$$_{CO}$$\it / (dv/dr)$ is assumed as $5 \times 10^{-5}$. 
  }\label{fig:LVG_6point}
\end{figure}

We selected six positions A-F for a LVG analysis in order to obtain temperature $T_{\rm k}$ and density as listed in Table 1. The six line profiles are shown in Figure\,\ref{fig:LVG_LineProfile}. Figure\,\ref{fig:LVG_6point} shows the plots of the curves of constant $^{12}$CO $J$=3--2/1--0 intensity ratio and the $^{13}$CO 1--0/$^{12}$CO 1--0 intensity ratio. The blue and red shaded areas show error limits in the two ratios. $X_{\rm CO}/dv/dr=5\times 10^{-5}$ is assumed. The solutions in each are shown in Table 1. The results show that $T_{\rm k}$ is 20--38\,K and density is in a range from $10^{3}$ to $4\times 10^{4}$\,cm$^{-3}$, confirming the high excitation conditions in the two clouds.  

The result of the analysis at six positions and a summary is given in Table 2. The blue-shifted cloud showing a high ratio with a lower bound of 20\,K. Considering that the other points in the blue-shifted cloud lie outside the intensity peak, it is likely that the blue-shifted cloud as a whole has Tk higher than 30\,K, supporting that the cloud is associated with the cluster. The red-shifted cloud shows high temperature of 30\,K, which decreases with the distance from the cluster. At point F separated from the cluster the red-shifted cloud shows $19^{+4}_{-3}$\,K, still significantly higher than 10\,K. We infer that the heating source is [DBS2003]179, and the red-shifted cloud is also associated with the cluster.

\clearpage

\section{Discussion}\label{sec:Dis}
We discuss the results obtained in the present study. First, we summarize the association of the two molecular clouds with the cluster [DBS2003]179, and examine a possibility that the two clouds collided with each other to trigger the formation of the cluster. 

\subsection{The two molecular clouds associated with [DBS2003]179}\label{sec:d1}
The present study revealed two molecular clouds toward the cluster [DBS2003]179, which have different velocities by $\sim$20\,km\,s$^{-1}$, and showed that they are physically associated with the cluster as proved by high $T_{\rm k}$ due to heating of the cluster. A large-scale CO distribution shows that the two clouds are extended along the Galactic plane from $347\fdg6$ to $347\fdg8$. The blue-shifted cloud is distributed in a latitude range from 0.15 to 0.3, and the red-shifted cloud is split into two latitude ranges from $0\fdg1$ to $0\fdg2$ and from $0\fdg2$ to $0\fdg3$, respectively (see Figure\,\ref{fig:Mopra13ASTE_LB}). The distribution of the line intensity ratios shows that the ratio tends to decrease with distance from the cluster as is consistent with the heating by the cluster. A LVG analysis shows high $T_{\rm k}$ above 20\,K, confirming radiative heating by the cluster. {\it Spitzer} 8\,$\mu$m shows good correspondence with the western edge of the blue-shifted cloud, lending support for the association with the cluster.

\begin{figure}[htbp]
  \begin{center}
  \includegraphics[width=14cm]{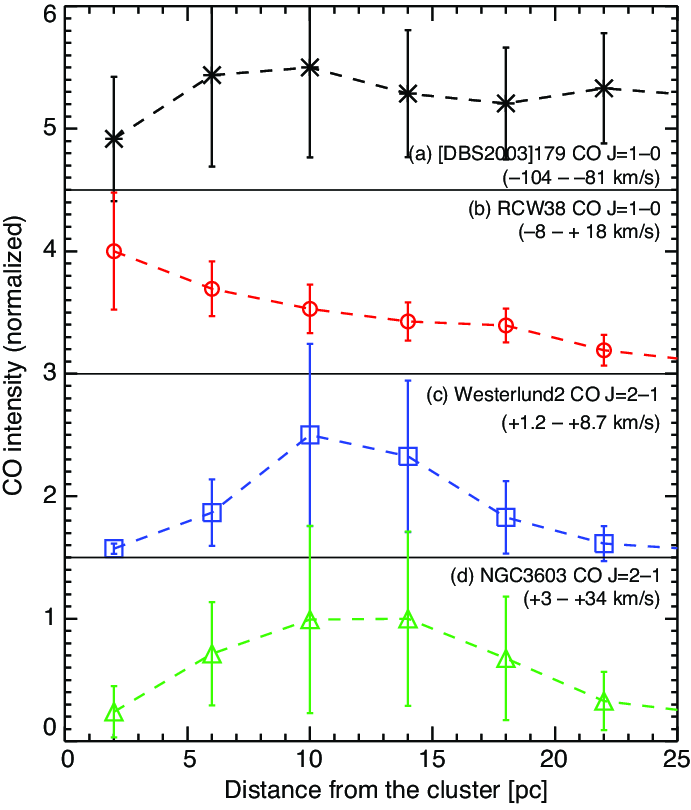}
  \end{center}
  \caption{The normalized $^{12}$CO ($J$=1--0) integrated intensity as a function of the distance to the cluster center for [DBS2003]179 (black), RCW38 (red), Westerlund 2 (blue), and NGC3603 (green), respectively. These calculations were conducted every 4\,pc.
  }\label{fig:radial_plot_NoMask}
\end{figure}

We examine the distribution of molecular gas with respect to the radius from the cluster center in [DBS2003]179. Figure\,\ref{fig:radial_plot_NoMask} shows a plot of the molecular cloud as a function of distance from the cluster center. The CO integrated intensity is averaged in annuli and normalized by the peak intensity. [DBS2003]179 shows similar trend with Westerlund 2 and NGC3603. The CO intensity is weak within 5\,pc of the center and increases to a peak value at 10\,pc, and then decreases with distance. In RCW38 the CO intensity shows the highest intensity toward the cluster and decreases outward. The difference among the clusters depends on the cluster age; [DBS2003]179, Westerlund 2, and NGC3603 have an age larger than 1--2\,Myr, whereas the age of RCW38 is 0.1\,Myr. In the youngest cluster RCW38 the dispersal of the parent cloud is least advanced. The cloud dispersal on 10\,pc scale proceeds rapidly from 0.1\,Myr to 1--2\,Myr.

\subsection{Distance of the cluster}\label{sec:d2}
The distance of the cluster is not settled in the literature. \cite{2012A&A...546A.110B} selected nine O stars in the cluster and derived a distance of $7.9 \pm 0.8$\,kpc by using Ks band data. \cite{2011AJ....142...40M} estimated a distance of 6.4\,kpc by using the Ks band data of a WR star MDM32 which was considered as a cluster member. \cite{2012MNRAS.419.1860D} derived a distance of 9.0\,kpc by assuming that the cluster belongs to molecular clouds near the direction of the cluster. The present study firmly identified the present two molecular clouds associated with the cluster, which are different from what \cite{2012MNRAS.419.1860D} adopted, and allows us to derive a kinematical distance more reliably. The distance of [DBS2003]179 is calculated to be $D = 5.26 \pm 0.18$\,kpc by using the flat rotation model toward $(l,b)(ep=J2000) = (347\fdg5764 , +00\fdg1865)$, where $R_0 = 8.34 \pm 0.16$\,kpc and $V_0 = 240 \pm 8$\,kms (\citet{2014ApJ...783..130R}) were adopted. The error in the kinematic distance however may not be small. Considering the above all we adopt the distance 6.4\,kpc (\cite{2011AJ....142...40M}) which is nearest to the present one in the literature. 

\begin{figure}[htbp]
  \begin{center}
  \includegraphics[width=14cm]{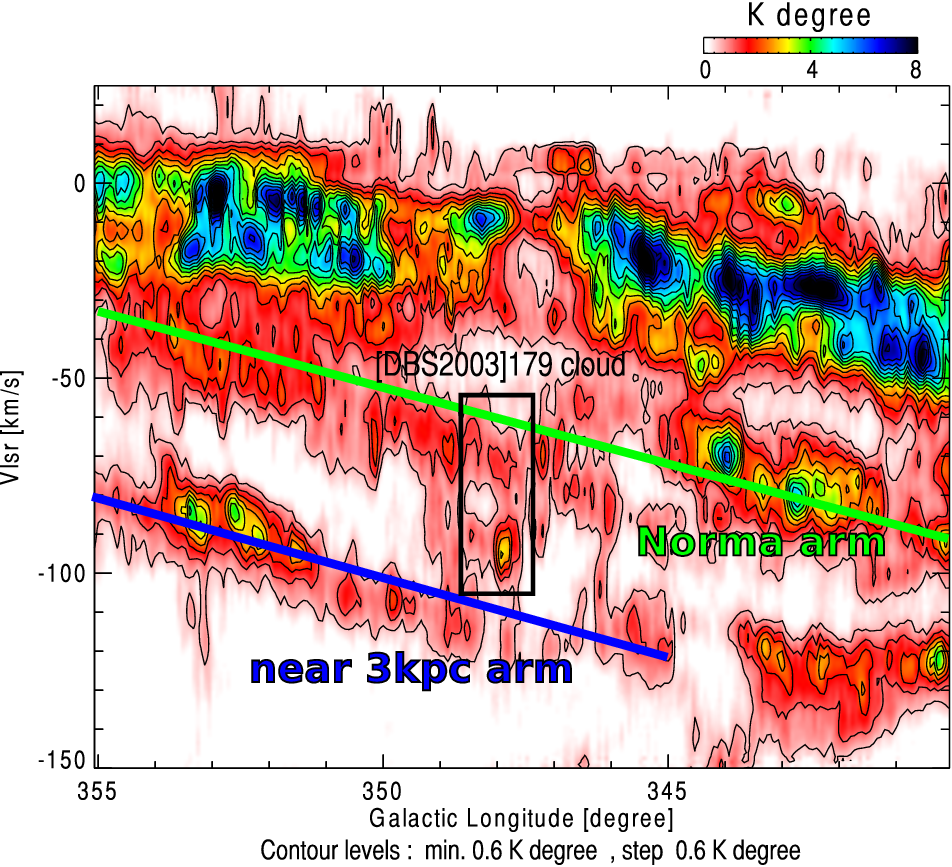}
  \end{center}
  \caption{The $^{12}$CO ($J$=1--0) $l$--$v$ diagram obtained with NANTEN2 taken from \citet{2010PASJ...62..557T} between the $b$ range of $-$1.0 and $+$0.93 degrees. The Norma arm and the near 3\,kpc arm are colored by green and blue, respectively (\cite{2011ApJ...733...27G,1993A&A...275...67B,2008ApJ...683L.143D}).
  }\label{fig:DBS_Norma}
\end{figure}

The red-shifted cloud at -75 \,km\,s$^{-1}$ shows continuous velocity distribution which is possibly tracing the spiral arm. Figure\,\ref{fig:DBS_Norma} shows a position--velocity diagram constructed base on the CO data in \cite{2010PASJ...62..557T}. The Figure\,\ref{fig:DBS_Norma} shows the Norma Arm as well as the 3\,kpc expanding Arm (\citet{2011ApJ...733...27G}, \citet{1993A&A...275...67B}, \citet{2008ApJ...683L.143D}). The present study showed that the red-shifted cloud coincides with the location of the Norma Arm in the longitude-velocity diagram. 

We calculate the column density of the clouds by using the Mopra $^{13}$CO ($J$=1--0) data and mass of the molecular clouds by using the NANTEN2 $^{12}$CO ($J$=1--0) data where $T_{\rm ex}$ of 30\,K is adopted. We adopt $N({\rm H_2})$\,[cm$^{-2}$] = $5.0\,\times\,10^5\,\times\,N(^{13}{\rm CO})$ for conversion of $N(^{13}$CO$)$ into NH$_2$ (\citet{1978ApJS...37..407D}) and an $X_{\rm CO}=2\times10^{20}$\,cm$^{-2}$\,(K \,km\,s$^{-1}$)$^{-1}$ (\citet{1988A&A...207....1S}) for conversion of W$^{12}$CO. As a result we estimate the column density and cloud mass of the blue-shifted cloud to be $8\times10^{22}$\,cm$^{-2}$ and the to be $2\times 10^5$\,$M_{\odot}$, respectively, and the column density of the red-shifted cloud and the column density and mass of the red-shifted cloud to be $5\times10^{22}$\,cm$^{-2}$ and $2\times 10^5$\,$M_{\odot}$, respectively. The masses are by an order of magnitude larger than the typical cloud masses associated with the {\it Spitzer} Bubble $10^4$\,$M_{\odot}$. The present mass and column density are similar to the cloud mass and the column density associated with the other super star clusters (Wd2 \citet{2009ApJ...696L.115F}; NGC3603 and RCW38 \citet{2014ApJ...780...36F}; \yearcite{2016ApJ...820...26F}).

\subsection{Triggered formation of the cluster by a cloud--cloud collision}\label{sec:d3}
The present results revealed that the two clouds show complementary distribution with each other (Figure\,\ref{fig:ASTEBridge}(a)) and that the bridge features are located in several places between the clouds (Figure\,\ref{fig:ASTEBridge}(b)). These two are typical signatures of colliding clouds as observed in young massive clusters and {\it Spitzer} bubble (e.g., \citet{2009ApJ...696L.115F}; \citet{2015ApJ...806....7T}). The velocity difference between the two clouds 20\,km\,s$^{-1}$, if we adopt a distance between the two cloud to be 30 pc, requires a total mass of $1.4\times10^6$\,$M_{\odot}$ for gravitationally binding. The mass is by an order of magnitude larger than the present cloud mass, and the clouds are not gravitationally bound, favoring a cloud--cloud collision by chance.

\begin{figure}[htbp]
  \begin{center}
  \includegraphics[width=14cm]{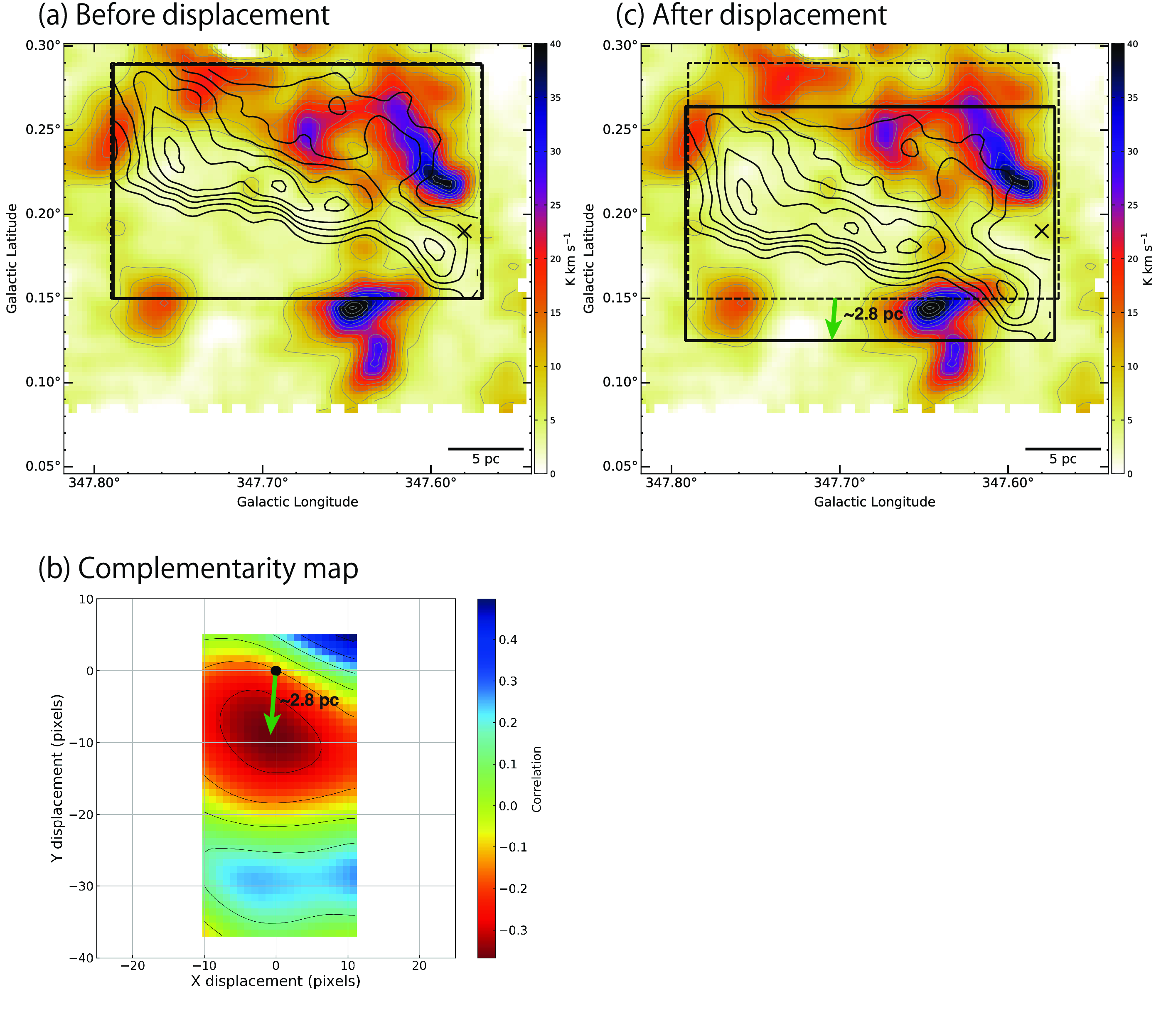}
  \end{center}
  \caption{(a) The $^{12}$CO ($J$=1--0) integrated intensity distributions of the blue-shifted cloud (color scale) and red-shifted cloud (contours). The contours are plotted at every 3.5\,K\,km\,s$^{-1}$ from 6.5\,K\,km\,s$^{-1}$. (b) The correlation coefficient distribution as a function of X-direction (Galactic Longitude) displacement and Y-direction (Galactic Latitude) displacement. The lower correlation coefficient (anti correlation) indicates that the complementarity is high. (c) The $^{12}$CO ($J$=1--0) integrated intensity distributions after the displacement.
  }\label{fig:disp}
\end{figure}

\begin{figure}[htbp]
  \begin{center}
  \includegraphics[width=14cm]{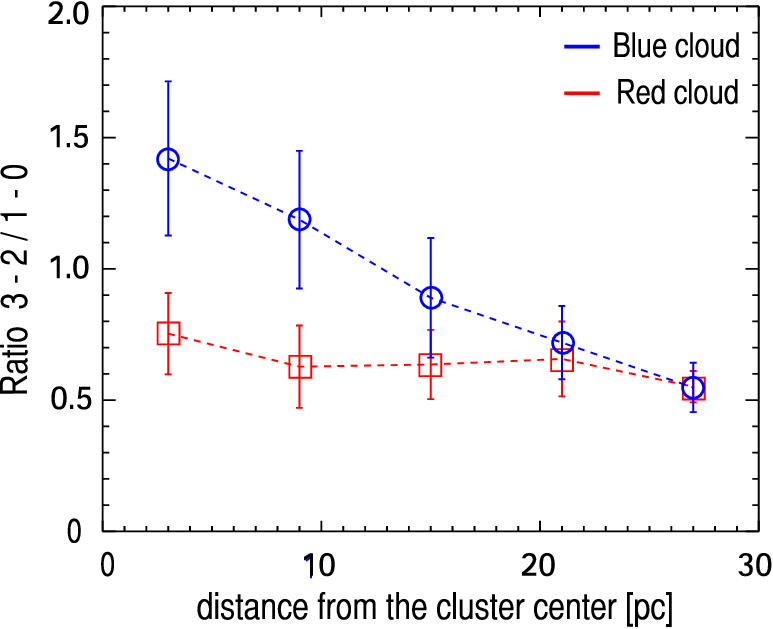}
  \end{center}
  \caption{The $^{12}$CO ($J$=3--2)/$^{12}$CO ($J$=1--0) integrated intensity ratio as a function of the distance to the cluster center for the blue-shifted cloud (blue lines) and red-shifted cloud (red lines). These calculations were conducted every 6\,pc.
  }\label{fig:Ratio-distance}
\end{figure}

We examine the formation process of the super star cluster [DBS2003]179 in terms of cloud--cloud collision. In order to estimate the collision time scale in the scenario, we divided a distance which the red-shifted cloud moved since the collision started by the velocity difference. We assumed that the relative position where the distributions of two clouds are most complementary corresponds to the cloud position when the collision was initiated. The most complementary position was calculated by using the same method in \cite{2018ApJ...859..166F}, but we used the Spearman's rank correlation coefficient between the integrated intensity distribution of the two clouds as the complementarity. Figure\,\ref{fig:disp}(a) shows the integrated intensity distribution of the two clouds. Figure\,\ref{fig:disp}(b) shows a correlation coefficient distribution as a function of X-direction (Galactic Longitude) displacement and Y-direction (Galactic Latitude) displacement. The lower correlation coefficient (anti correlation) indicates that the complementarity is high. As a result, the distance travelled by the red-shifted cloud on the sky plane was estimated as $\sim$2.8\,pc in this method, and the cloud distributions after displacement are shown in Figure\,\ref{fig:disp}(c). If we tentatively assume that the relative velocity along the direction of the displacement is the same as the relative radial velocity of $\sim$20 \,km\,s$^{-1}$, a collision timescale is estimated to be $\sim$2.8 pc/$\sim$20 \,km\,s$^{-1}$ = 0.1--0.2\,Myr. On the other hand, the blue-shifted cloud shows intensity depression within 10\,pc from the cluster position (Figure\,\ref{fig:Mopra13ASTE_LB}), while the red-shifted cloud showing no hint of intensity depression by the ionization, suggesting that the cloud is separated from the cluster by more than 10 pc. This poses a separation of the red-shifted cloud from the cluster to be more than 10 pc. If we assume that the ionization proceeds at 5 \,km\,s$^{-1}$ (e.g., \citet{2016ApJ...820...26F}), the ionization time scale becomes 2\,Myr. Figure\,\ref{fig:Ratio-distance} shows a plot of the ratio as a function of radius from the cluster center. Since the two curves become close with each other beyond 20\,pc from the cluster, we infer that the red-shifted cloud is located at a distance of more than 20\,pc along the line of sight. We estimate the collision timescale to be longer than 1\,Myr from a ratio of the velocity and the distance, 20 pc/20\,km\,s$^{-1}$. Considering these above, we conservatively estimate the collision time scale to be 1-2\,Myr, and the relative velocity along the direction of the displacement seems to be small (a few \,km\,s$^{-1}$). This is consistent with a cluster age of 2-5\,Myr estimated by \citet{2008A&A...488..151B}. \citet{2012A&A...546A.110B} estimated the mass of an OIf star to be 40--80\,$M_{\odot}$.  \citet{2012ApJ...750L..44K} derived a duration of star formation is very short in the order of 0.1Myr in NGC3603 and Wd1. This indicates that the cluster members of a young massive cluster are formed simultaneously in a short time span as compared with the cluster age. This is consistent with that the age of high mass stars are in a small range $\sim$0.7\,Myr (\citet{2012A&A...546A.110B}) . 

DBS2003[179] includes three WR stars and at least 10 O stars. There is a possibility that the stellar winds of [DBS2003]179 accelerated gas to cause a velocity separation of 20\,km\,s$^{-1}$. In order to examine this feedback, we compare the energy available to accelerate the gas in Wd2 $3.6\times 10^{51}$\,erg (\cite{2007A&A...463..981R}) and that of NGC3603 $5\times 10^{51}$\,erg (\citet{2008ApJ...675.1319H}). The kinetic energy of molecular gas of $4\times 10^5$\,$M_{\odot}$ having velocity of 20\,km\,s$^{-1}$ is $\sim$ $1\times 10^{51}$\,erg. By assuming that wind energy similar to these is available in [DBS2003]179, the kinetic energy corresponds to $\sim$20\% of the stellar wind energy, $(4-5)\times 10^{51}$\,erg. \cite{1977ApJ...218..377W} argued that 20\% of the stellar wind energy is likely converted to the expansion energy of neutral gas under the adiabatic condition. \citet{2008IAUS..250..355A} however argued that only a few \% of the kinetic energy in stellar winds can be converted into the kinetic energy of the neutral gas. It seems difficult to accelerate gas far from the cluster by more than 10\,pc. In case of acceleration by the winds, we expect a velocity gradient in the sense that the acceleration is larger toward the cluster, whereas such a velocity gradient is not seen (the moment 1 is shown in Figure\,\ref{fig:moms}). Accordingly, we infer that the stellar winds do not play a role in accelerating gas in [DBS2003]179. The same conclusion was reached in Wd2 and a cloud--cloud collision was concluded by \cite{2009ApJ...696L.115F}.

In the cloud--cloud collision scenario, it is argued that the encounter between the two clouds is by chance. Such relative motion arises in the Galactic disk as a cumulative effects of ISM acceleration by supernova explosions, stellar winds, in addition to the disk gravity including the spiral arms, and the two clouds behave randomly. It is not required to introduce an external force to accelerate the clouds.  

To summarize, we discovered two molecular clouds are associated with [DBS2003]179, and that the two cloud show two signatures typical to cloud--cloud collision, i.e., the complementary distribution and the bridge features between the two clouds. The time scale of the collision is estimated to be consistent with the cluster age, supporting that the cluster formation was triggered by the collision between the two clouds. We frame a scenario that the cluster [DBS2003]179 was formed by a trigger of a cloud--cloud collision; 1--2\,Myr ago the red-shifted cloud at -95\,km\,s$^{-1}$ collided by chance with the blue-shifted cloud at -75\,km\,s$^{-1}$. The collision created a compressed layer between the two clouds where a high-density turbulent layer were produced and multiple O stars were formed as a cluster. The collision produced the complementary distribution between the two clouds. The UV radiation emitted by the cluster ionized the inner $\sim$5\,pc of the cluster, creating thereby the cavity of molecular gas around the cluster. The blue-shifted cloud shows strong effects due to the heating and ionization because the distance of the cluster is closer to the cluster than the red-shifted cloud as indicated by the higher line intensity ratio in the blue-shifted cloud than in the red-shifted cloud.

\subsection{Comparison with the other super star clusters}\label{sec:d4}
We discuss the common properties of young super star clusters studied so far.
\cite{2016ApJ...820...26F} presented discussion on the three young super star clusters and a small cluster M20 and a {\it Spitzer} bubble RCW120, where cloud--cloud collisions are shown to be taking place. Table3 lists observed physical properties of the four super star clusters, Westerlund 2 , NGC3603 , RCW38 , [DBS2003]179. The typical mass of the molecular clouds is greater than $10^{4}$\,$M_{\odot}$ and high molecular column density $10^{22}$\,cm$^{-2}$--$10^{23}$\,cm$^{-2}$. This indicates that it is not necessary for both of the colliding clouds are massive and dense. If one of them satisfies the condition, formation of high mass star/ clusters can be triggered. Conversely, single O star formation takes place for mass and column density by a factor of ten smaller than in the {\it Spitzer} Bubble (RCW120) and smaller cluster having an O star.

\section{Conclusions}
We carried out a study of the molecular clouds toward a super star cluster [DBS2003]179 by using multi-transitions of CO with NANTEN2, Mopra and ASTE. The study has revealed the following properties of the CO clouds toward the cluster;

\begin{enumerate}
\item We found two molecular clouds at -95\,km\,s$^{-1}$ (the blue-shifted cloud) and -75\,km\,s$^{-1}$ (the red-shifted cloud), both of which are elongated along the Galactic plane. The blue-shifted cloud has high column density of $5\times 10^{22}$\,cm$^{-2}$, while the red-shifted cloud low column density of $8\times 10^{22}$\,cm$^{-2}$. The clouds have molecular mass of $2\times 10^5$\,$M_{\odot}$ and $2\times 10^5$\,$M_{\odot}$, respectively, at a distance of 6\,kpc.

\item The two clouds are physical associated with each other as well as with the cluster by the following reasons; a) The red-shifted cloud show clear correlation with the {\it Spitzer} 8\,$\mu$m emission, showing signatures of heating by the cluster, b) The blue-shifted cloud shows much weaker 8\,$\mu$m emission, whereas a small peak corresponding to the peak exists, indicating that the cloud is moderately heated by the cluster, c) Kinematically, the two clouds are linked by bridge features at several places, and 4) the two clouds show spatially complementary distribution with each other. 

\item The distribution of an intensity ratio $^{12}$CO ($J$=3--2) / $^{12}$CO ($J$=1--0) (R3--2/1--0) indicates that most of the blue-shifted cloud shows R3--2/1--0 of $\sim$1.0. Especially toward the region of bright {\it Spitzer} 8$\mu$m emission R3--2/1--0 becomes as high as $\sim$1.5. The red-shifted cloud shows R3--2/1--0 of $\sim$0.8 with no strong variation while it decreases gradually with distance from the cluster center. The ratio more than 0.8 is significantly higher than the typical clouds with no extra heating (ref), indicating significant radiative heating by the cluster.

\item An LVG analysis of the three transitions $^{12}$CO ($J$=3--2), ($J$=1--0) and $^{13}$ ($J$=1--0) shows that the kinetic temperature of the two clouds is in a range from 20\,K to more than 50\,K, lending support for the radiative heating and their physical association with the cluster. These temperatures are significantly higher than 10\,K in a typical Galactic cloud without extra heating.

\item We frame a hypothesis that the two clouds collided with each other, with a typical time scale of 1--2\,Myr. The collision compressed the interface layer between the two clouds and triggered the formation of the super star cluster. Each of the two clouds has mass of $10^5$\,$M_{\odot}$ at the distance.

\item We examined an alternative scenario that the external effects on the origin of the velocity difference between the two clouds identified in the present study. We considered SNRs, stellar winds, and cloud--cloud collision as the effects. As a result, the former two effects are excluded based on quantitative considerations, and a cloud--cloud collision is found to be a viable mechanism. We presented that the cluster age estimated from the cloud--cloud collision is consistent within a factor two with the age of the cluster derived in the previous works. 

\item We compared the physical parameters of the cloud--cloud collisions and found that one of the clouds in collision is required to have column density of $10^{23}$\,cm$^{-2}$ and mass of $10^5$\,$M_{\odot}$ in order to form a young massive cluster with more than 10 O stars. This lends support for the conclusion reached by \cite{2018ApJ...859..166F}.

\end{enumerate}

\clearpage

\begin{table}
  \tbl{Young Massive Clusters in the Galaxy}{%
  \begin{tabular}{lcccccccccl}
   \hline \hline
    Name  & \it l  & \it b & \it D & Age & $\log (M_{clus} /  M_{\odot} )$  & Radius & Molecular clouds & References$^{a}$ & References$^{b}$    \\
     & [deg] & [deg] & [kpc] & [Myr] & & [pc] & & &  \\
    \hline
    Arches & 0.12 & 0.02 & 8.0 & 2.0 & 4.3 & 0.4 & No & [1] &  \\
    Quintuplet & 0.16 & -0.06 & 8.2 & 4.0 & 4.0 & 2.0 & No & [1] &  \\
    RCW38 & 268.03 & -0.98 & 1.7 & 0.5 & … & 0.8 & Yes & [2] & [10]  \\
    Westerlund 2 & 284.25 & -0.40 & 5.4 & 2.0 & 4.0 & 0.8 & Yes & [3] & [11,12]  \\
    Trumpler 14 & 287.41 & -0.58  & 2.6 & 2.0 & 4.0 & 0.5 & No & [4] &  \\
    NGC 3603 & 291.62 & -0.52 & 7.0 & 2.0 & 4.1 & 0.7 & Yes & [5] & [13]  \\
    Westerlund 1 & 339.55 & -0.40 & 5.2 & 3.5 & 4.5 & 1.0 & No & [6] &  \\
    $[$DBS2003]179 & 347.58 & 0.19 & 6.4 & 2~5 & 4.4 & 0.5 & Yes & [7,8,9] & This work  \\
    \hline
  \end{tabular}}\label{table1.1}
  \begin{tabnote}
    Column (1) Name of the clusters (2),(3) Position of the clusters (4) Distance to the clusters (5) Age of the clusters (6) Stellar mass of the clusters (7) Radius of the clusters (8) Associated molecular clouds (Yes/No) *References: [1] \cite{1999ApJ...525..750F}; [2] \cite{2006AJ....132.1100W}; [3] \cite{2009A&A...498L..37P}; [4] \cite{2007A&A...476..199A};  [5] \cite{2008ApJ...675.1319H}; [6] \cite{2005A&A...434..949C}; [7] \cite{2012A&A...546A.110B}; [8] \cite{2008A&A...488..151B}; [9] \cite{2011AJ....142...40M};  [10] \cite{2016ApJ...820...26F}; [11] \cite{2009ApJ...696L.115F}; [12] \cite{2010ApJ...709..975O}; [13] \cite{2014ApJ...780...36F}.
  \end{tabnote} 
\end{table}

\begin{table}
  \tbl{Results of the LVG analysis}{%
  \begin{tabular}{lcccccccc}
  \hline \hline
    Name  & position & \it l  & \it b & V$_{lsr}$ & Ratio$_{1}$ & Ratio$_{2}$ & T$_{kin}$ & n(H$_{2}$)    \\
     &  & [deg] & [deg] & [km\,s$^{-1}]$ & &  & [K] & [cm$^{-3}$]  \\
     (1) & (2) & (3) & (4) & (5) & (6) & (7) & (8) & (9) \\
    \hline
   Blue component & A & 347.62 & 0.26 & -97, -90 & 0.83 & 0.30 & $ 30^{+20}_{-10}$ & $3.4^{+5.6}_{-1.4} \times 10^{3}$  \\ 
   & B & 347.62 & 0.24 & -98, -88 & 0.95 & 0.20 & $\ge 109_{-54}$ & $\ge 3.3_{-3.1} \times 10^{4}$ \\ 
   & C & 347.60 & 0.14 & -97, -87 & 0.73 & 0.14 & $ 38^{+28}_{-13}$ & $ 1.2^{+1.1}_{-0.2} \times 10^{3}$ \\ 
   Red component & D & 347.63 & 0.21 & -80, -70 & 0.76 & 0.18 & $33^{+7}_{-6}$ & $1.6^{+0.7}_{-0.6} \times 10^{3}$ \\ 
   & E & 347.71 & 0.22 & -77, -66 & 0.73 & 0.26 & $23^{+6}_{-4}$ & $1.8^{+0.5}_{-0.3} \times 10^{3}$ \\ 
   & F & 347.78 & 0.24 & -76, -70 & 0.54 & 0.18 & $19^{+4}_{-3}$ & $9.1^{+1.0}_{-1.5} \times 10^{2}$ \\ 
    \hline
  \end{tabular}}\label{table3.4.1}
  \begin{tabnote}
    Column  (2) Point (Figure\,\ref{fig:Ratio3-2.1-0}) (3),(4) Position (5) Integrated velocity width (6) $^{12}$CO $J$=3--2/$J$=1--0 intensity ratio (7) $^{13}$CO/$^{12}$CO $J$=1--0 intensity ratio (8) Kinetic tempature (9) Volume density of H$_2$ 
    \end{tabnote} 
\end{table}

\begin{table}
  \tbl{Comparison of the six of cloud--cloud collision regions}{%
  \begin{tabular}{lccccc}
     \hline \hline
    Name  & cloud mass & column densities & velocity separation & created O stars & references    \\
     & [$M_{\odot}]$ & [cm$^{-2}]$ & [km\,s$^{-1}]$ &  &   \\
     (1) & (2) & (3) & (4) & (5) & (6) \\
    \hline
    $[$DBS2003]179 & 2$\times$10$^{5}$, 2$\times$10$^{5}$ & 8$\times$10$^{22}$,5$\times$10$^{22}$ & 20 & $\ge 10$ & This work  \\
    RCW38 & 2$\times$10$^{4}$, 3$\times$10$^{3}$ & 1$\times$10$^{23}$,1$\times$10$^{22}$ & 12 & $\sim$20 & [1]  \\
    NGC 3603 & 7$\times$10$^{4}$, 1$\times$10$^{4}$ & 1$\times$10$^{23}$,1$\times$10$^{22}$ & 20 & $\sim$30 & [2] \\
    Westerlund 2 & 8$\times$10$^{4}$, 9$\times$10$^{4}$ & 2$\times$10$^{23}$,2$\times$10$^{22}$ & 13 & 14 & [3,4]   \\
    M20 & 1$\times$10$^{3}$, 1$\times$10$^{3}$ & 1$\times$10$^{22}$,1$\times$10$^{22}$ & 7 & 1 & [5,6]   \\  
    RCW120 & 4$\times$10$^{3}$, 5$\times$10$^{4}$ & 8$\times$10$^{21}$,3$\times$10$^{22}$ & 20 & 1 & [7]   \\  
    \hline
  \end{tabular}}\label{table4.4}
  \begin{tabnote}
    Column (1) Name of the clusters (2),(3) Molecular mass and column density (Left: blue-shifted cloud, Right: red-shifted cloud) (4) Velocity separation (5) Number of O stars *References: [1] \cite{2016ApJ...820...26F},  [2] \cite{2014ApJ...780...36F}, [3] \cite{2009ApJ...696L.115F}, [4] \cite{2010ApJ...709..975O}, [5] \cite{2011ApJ...738...46T}, [6] \cite{2017ApJ...835..142T}, [7] \cite{2015ApJ...806....7T}.
  \end{tabnote} 
\end{table}

\clearpage

\begin{ack}
This study was financially supported by Grants-in-Aid for Scientific Research (KAKENHI) of the Japanese Society for the Promotion of Science (JSPS, grant numbers 15H05694 and 15K17607). The authors grateful to the NANTEN team and also would like to thank NASA, National Radio Astronomy Observatory (NRAO), and National Astronomy Observatory of Japan (NAOJ) for providing the data of Spitzer Space Telescope and ASTE.
\end{ack}

\clearpage



\begin{thebibliography}{}
\bibitem[Ascenso et al.(2007)]{2007A&A...476..199A} Ascenso, J., Alves, J., Vicente, S., et al.\ 2007, \aap, 476, 199
\bibitem[Arp, \& Sandage(1985)]{1985AJ.....90.1163A} Arp, H., \& Sandage, A.\ 1985, \aj, 90, 1163
\bibitem[Arthur(2008)]{2008IAUS..250..355A} Arthur, S.~J.\ 2008, Massive Stars as Cosmic Engines, 355
\bibitem[Borissova et al.(2008)]{2008A&A...488..151B} Borissova, J., Ivanov, V.~D., Hanson, M.~M., et al.\ 2008, \aap, 488, 151
\bibitem[Borissova et al.(2012)]{2012A&A...546A.110B} Borissova, J., Georgiev, L., Hanson, M.~M., et al.\ 2012, \aap, 546, A110
\bibitem[Brand, \& Blitz(1993)]{1993A&A...275...67B} Brand, J., \& Blitz, L.\ 1993, \aap, 275, 67
\bibitem[Caswell, \& Haynes(1987)]{1987A&A...171..261C} Caswell, J.~L., \& Haynes, R.~F.\ 1987, \aap, 171, 261
\bibitem[Clark et al.(2005)]{2005A&A...434..949C} Clark, J.~S., Negueruela, I., Crowther, P.~A., et al.\ 2005, \aap, 434, 949
\bibitem[Cohen et al.(2011)]{2011MNRAS.415.3354C} Cohen, D.~H., Gagn{\'e}, M., Leutenegger, M.~A., et al.\ 2011, \mnras, 415, 3354
\bibitem[Crowther et al.(2010)]{2010MNRAS.408..731C} Crowther, P.~A., Schnurr, O., Hirschi, R., et al.\ 2010, \mnras, 408, 731
\bibitem[Dame, \& Thaddeus(2008)]{2008ApJ...683L.143D} Dame, T.~M., \& Thaddeus, P.\ 2008, \apjl, 683, L143
\bibitem[Davies et al.(2012)]{2012MNRAS.419.1860D} Davies, B., de La Fuente, D., Najarro, F., et al.\ 2012, \mnras, 419, 1860
\bibitem[Dickman(1978)]{1978ApJS...37..407D} Dickman, R.~L.\ 1978, \apjs, 37, 407
\bibitem[Dutra et al.(2003)]{2003A&A...400..533D} Dutra, C.~M., Bica, E., Soares, J., et al.\ 2003, \aap, 400, 533
\bibitem[Elmegreen(1998)]{1998ASPC..148..150E} Elmegreen, B.~G.\ 1998, Origins, 150
\bibitem[Evans et al.(2010)]{2010ApJS..189...37E} Evans, I.~N., Primini, F.~A., Glotfelty, K.~J., et al.\ 2010, \apjs, 189, 37
\bibitem[Ezawa et al.(2004)]{2004SPIE.5489..763E} Ezawa, H., Kawabe, R., Kohno, K., et al.\ 2004, \procspie, 763
\bibitem[Ezawa et al.(2008)]{2008SPIE.7012E..08E} Ezawa, H., Kohno, K., Kawabe, R., et al.\ 2008, \procspie, 701208
\bibitem[Figer et al.(1999)]{1999ApJ...525..750F} Figer, D.~F., Kim, S.~S., Morris, M., et al.\ 1999, \apj, 525, 750
\bibitem[Frerking et al.(1982)]{1982ApJ...262..590F} Frerking, M.~A., Langer, W.~D., \& Wilson, R.~W.\ 1982, \apj, 262, 590
\bibitem[Fukui et al.(2014)]{2014ApJ...780...36F} Fukui, Y., Ohama, A., Hanaoka, N., et al.\ 2014, \apj, 780, 36
\bibitem[Fukui et al.(2016)]{2016ApJ...820...26F} Fukui, Y., Torii, K., Ohama, A., et al.\ 2016, \apj, 820, 26
\bibitem[Fukui et al.(2018)]{2018ApJ...859..166F} Fukui, Y., Torii, K., Hattori, Y., et al.\ 2018, \apj, 859, 166
\bibitem[Furukawa et al.(2009)]{2009ApJ...696L.115F} Furukawa, N., Dawson, J.~R., Ohama, A., et al.\ 2009, \apjl, 696, L115
\bibitem[Goldreich, \& Kwan(1974)]{1974ApJ...189..441G} Goldreich, P., \& Kwan, J.\ 1974, \apj, 189, 441
\bibitem[Green et al.(2011)]{2011ApJ...733...27G} Green, J.~A., Caswell, J.~L., McClure-Griffiths, N.~M., et al.\ 2011, \apj, 733, 27
\bibitem[Hands et al.(2004)]{2004yCat..73510031H} Hands, A.~D.~P., Warwick, R.~S., Watson, M.~G., et al.\ 2004, VizieR Online Data Catalog, J/MNRAS/351/31
\bibitem[Harayama et al.(2008)]{2008ApJ...675.1319H} Harayama, Y., Eisenhauer, F., \& Martins, F.\ 2008, \apj, 675, 1319
\bibitem[Inoue et al.(2008)]{2008stt..conf..281I} Inoue, H., Muraoka, K., Sakai, T., et al.\ 2008, Ninteenth International Symposium on Space Terahertz Technology, 281
\bibitem[Johnson(2005)]{2005IAUS..227..413J} Johnson, K.~E.\ 2005, Massive Star Birth: A Crossroads of Astrophysics, 413
\bibitem[Kohno et al.(2004)]{2004dimg.conf..349K} Kohno, K., Yamamoto, S., Kawabe, R., et al.\ 2004, The Dense Interstellar Medium in Galaxies, 349
\bibitem[Kudryavtseva et al.(2012)]{2012ApJ...750L..44K} Kudryavtseva, N., Brandner, W., Gennaro, M., et al.\ 2012, \apjl, 750, L44
\bibitem[Ladd et al.(2005)]{2005PASA...22...62L} Ladd, N., Purcell, C., Wong, T., et al.\ 2005, PASA, 22, 62
\bibitem[Leung et al.(1984)]{1984ApJS...56..231L} Leung, C.~M., Herbst, E., \& Huebner, W.~F.\ 1984, \apjs, 56, 231
\bibitem[Mangum et al.(2007)]{2007A&A...474..679M} Mangum, J.~G., Emerson, D.~T., \& Greisen, E.~W.\ 2007, \aap, 474, 679
\bibitem[Mauerhan et al.(2011)]{2011AJ....142...40M} Mauerhan, J.~C., Van Dyk, S.~D., \& Morris, P.~W.\ 2011, \aj, 142, 40
\bibitem[Ohama et al.(2010)]{2010ApJ...709..975O} Ohama, A., Dawson, J.~R., Furukawa, N., et al.\ 2010, \apj, 709, 975
\bibitem[Oka et al.(2007)]{2007PASJ...59...15O} Oka, T., Nagai, M., Kamegai, K., et al.\ 2007, \pasj, 59, 15
\bibitem[Pfalzner(2009)]{2009A&A...498L..37P} Pfalzner, S.\ 2009, \aap, 498, L37
\bibitem[Pfeffermann and Aschenbach(1996)]{1996rftu.proc..267P} Pfeffermann, E., Aschenbach, B.\ 1996.\ ROSAT observation of a new supernova remnant in the constellation Scorpius..\ Roentgenstrahlung from the Universe 267.
\bibitem[Portegies Zwart et al.(2010)]{2010ARA&A..48..431P} Portegies Zwart, S.~F., McMillan, S.~L.~W., \& Gieles, M.\ 2010, \araa, 48, 431
\bibitem[Rodgers et al.(1960)]{1960MNRAS.121..103R} Rodgers, A.~W., Campbell, C.~T., \& Whiteoak, J.~B.\ 1960, \mnras, 121, 103
\bibitem[Sano et al.(2010)]{2010ApJ...724...59S} Sano, H., Sato, J., Horachi, H., et al.\ 2010, \apj, 724, 59
\bibitem[Sano et al.(2013)]{2013ApJ...778...59S} Sano, H., Tanaka, T., Torii, K., et al.\ 2013, \apj, 778, 59
\bibitem[Sawada et al.(2008)]{2008PASJ...60..445S} Sawada, T., Ikeda, N., Sunada, K., et al.\ 2008, \pasj, 60, 445
\bibitem[Sorai et al.(2000)]{2000SPIE.4015...86S} Sorai, K., Sunada, K., Okumura, S.~K., et al.\ 2000, \procspie, 86
\bibitem[Strong et al.(1988)]{1988A&A...207....1S} Strong, A.~W., Bloemen, J.~B.~G.~M., Dame, T.~M., et al.\ 1988, \aap, 207, 1
\bibitem[Rauw et al.(2007)]{2007A&A...463..981R} Rauw, G., Manfroid, J., Gosset, E., et al.\ 2007, \aap, 463, 981
\bibitem[Reid et al.(2014)]{2014ApJ...783..130R} Reid, M.~J., Menten, K.~M., Brunthaler, A., et al.\ 2014, \apj, 783, 130
\bibitem[Takeuchi et al.(2010)]{2010PASJ...62..557T} Takeuchi, T., Yamamoto, H., Torii, K., et al.\ 2010, \pasj, 62, 557
\bibitem[Tan et al.(2014)]{2014prpl.conf..149T} Tan, J.~C., Beltr{\'a}n, M.~T., Caselli, P., et al.\ 2014, Protostars and Planets VI, 149
\bibitem[Torii et al.(2011)]{2011ApJ...738...46T} Torii, K., Enokiya, R., Sano, H., et al.\ 2011, \apj, 738, 46
\bibitem[Torii et al.(2015)]{2015ApJ...806....7T} Torii, K., Hasegawa, K., Hattori, Y., et al.\ 2015, \apj, 806, 7
\bibitem[Torii et al.(2017)]{2017ApJ...835..142T} Torii, K., Hattori, Y., Hasegawa, K., et al.\ 2017, \apj, 835, 142
\bibitem[Wang et al.(1994)]{1994ApJS...95..503W} Wang, Y., Jaffe, D.~T., Graf, U.~U., et al.\ 1994, \apjs, 95, 503
\bibitem[Weaver et al.(1977)]{1977ApJ...218..377W} Weaver, R., McCray, R., Castor, J., et al.\ 1977, \apj, 218, 377
\bibitem[Wilson et al.(1970)]{1970A&A.....6..364W} Wilson, T.~L., Mezger, P.~G., Gardner, F.~F., et al.\ 1970, \aap, 6, 364
\bibitem[Wilson, \& Rood(1994)]{1994ARA&A..32..191W} Wilson, T.~L., \& Rood, R.\ 1994, \araa, 32, 191
\bibitem[Wolfire, \& Cassinelli(1987)]{1987ApJ...319..850W} Wolfire, M.~G., \& Cassinelli, J.~P.\ 1987, \apj, 319, 850
\bibitem[Wolk et al.(2006)]{2006AJ....132.1100W} Wolk, S.~J., Spitzbart, B.~D., Bourke, T.~L., et al.\ 2006, \aj, 132, 1100
\bibitem[Zinnecker, \& Yorke(2007)]{2007ARA&A..45..481Z} Zinnecker, H., \& Yorke, H.~W.\ 2007, \araa, 45, 481
\end{thebibliography}
\end{document}